\documentclass[12pt]{article}

\usepackage{amsmath,amsthm, amsfonts, amssymb, amsxtra,amsopn}
\usepackage{pgfplots}
\usepgfplotslibrary{colormaps}
\pgfplotsset{compat=1.15}
\usepackage{pgfplotstable}
\usetikzlibrary{pgfplots.statistics} 
\usepackage{filecontents}
\usepackage{graphicx}
\usepackage{multirow}
\usepackage{tabu}
\usepackage{algorithmic}
\usepackage{longtable}
\usepackage{booktabs}
\usepackage{listings}
\usepackage{cmap}
\usepackage{colortbl}
\usepackage{adjustbox}
\usepackage{epsfig}
\usepackage{xstring}
\usepackage{subfigure}

\PassOptionsToPackage{hyphens}{url}

\usepackage{hyperref}
\hypersetup{colorlinks=true,linkcolor=black,citecolor=black,urlcolor=blue,filecolor=black}
\hypersetup{pdfpagemode=UseNone,pdfstartview=}

\usepackage[tableposition=top]{caption}
\usepackage{enumitem}
\setlist[itemize]{noitemsep, topsep=0pt}

\usepackage{array}
\newcolumntype{C}[1]{>{\centering\let\newline\\\arraybackslash\hspace{0pt}}m{#1}}

\advance\oddsidemargin by -0.45in
\advance\textwidth by 0.9in

\advance\topmargin by -0.3in
\advance\textheight by 0.6in

\long\def\symbolfootnotetext[#1]#2{\begingroup%
\def\thefootnote{\fnsymbol{footnote}}\footnotetext[#1]{#2}\endgroup}

\DeclareMathOperator{\stst}{st}
\DeclareMathOperator{\ndnd}{nd}
\DeclareMathOperator{\rdrd}{rd}
\DeclareMathOperator{\thth}{th}

\def\z{\phantom{0}}

\pgfkeys{
    /pgf/number format/fixed zerofill=true }
    
\pgfplotstableset{
    color cells/.code={%
        \pgfqkeys{/color cells}{#1}%
        \pgfkeysalso{%
            postproc cell content/.code={%
                \begingroup
                \pgfkeysgetvalue{/pgfplots/table/@preprocessed cell content}\value
\ifx\value\empty
\endgroup
\else
                \pgfmathfloatparsenumber{\value}%
                \pgfmathfloattofixed{\pgfmathresult}%
                \let\value=\pgfmathresult
                \pgfplotscolormapaccess
                    [\pgfkeysvalueof{/color cells/min}:\pgfkeysvalueof{/color cells/max}]%
                    {\value}%
                    {\pgfkeysvalueof{/pgfplots/colormap name}}%
                \pgfkeysgetvalue{/pgfplots/table/@cell content}\typesetvalue
                \pgfkeysgetvalue{/color cells/textcolor}\textcolorvalue
                \toks0=\expandafter{\typesetvalue}%
                \xdef\temp{%
                    \noexpand\pgfkeysalso{%
                        @cell content={%
                            \noexpand\cellcolor[rgb]{\pgfmathresult}%
                            \noexpand\definecolor{mapped color}{rgb}{\pgfmathresult}%
                            \ifx\textcolorvalue\empty
                            \else
                                \noexpand\color{\textcolorvalue}%
                            \fi
                            \the\toks0 %
                        }%
                    }%
                }%
                \endgroup
                \temp
\fi
            }%
        }%
    }
}

\def\srowvecc#1#2{(\!\begin{array}{cc} 
      \noexpandarg\IfBeginWith{#1}{-}{\! #1}{#1}
    & #2\kern-0.5pt\end{array}\!)}
\def\rowvecc#1#2{\left(\!\begin{array}{cc} 
      \noexpandarg\IfBeginWith{#1}{-}{\! #1}{#1}
    & #2\kern-0.5pt\end{array}\!\right)}
\def\rowveccc#1#2#3{\left(\!\begin{array}{ccc} 
      \noexpandarg\IfBeginWith{#1}{-}{\! #1}{#1}
    & #2 
    & #3\kern-0.5pt\end{array}\!\right)}
\def\rowvecccc#1#2#3#4{\left(\!\begin{array}{cccc}
      \noexpandarg\IfBeginWith{#1}{-}{\! #1}{#1}
    & #2 
    & #3 
    & #4\kern-0.5pt\end{array}\!\right)}
\def\srowvecccc#1#2#3#4{\bigl(\!\begin{array}{cccc}
      \noexpandarg\IfBeginWith{#1}{-}{\! #1}{#1}
    & #2 
    & #3 
    & #4\kern-0.5pt\end{array}\!\bigr)}
\def\rowveccccc#1#2#3#4#5{\left(\!\begin{array}{ccccc} 
      \noexpandarg\IfBeginWith{#1}{-}{\! #1}{#1}
    & #2
    & #3
    & #4
    & #5\kern-0.5pt\end{array}\!\right)}
\def\srowvecccccc#1#2#3#4#5#6{(\!\begin{array}{cccccc} 
      \noexpandarg\IfBeginWith{#1}{-}{\! #1}{#1}
    & #2
    & #3
    & #4
    & #5
    & #6\kern-0.5pt\end{array}\!)}
\def\rowvecccccc#1#2#3#4#5#6{\left(\!\begin{array}{cccccc} 
      \noexpandarg\IfBeginWith{#1}{-}{\! #1}{#1}
    & #2
    & #3
    & #4
    & #5
    & #6\kern-0.5pt\end{array}\!\right)}

\title{Auxiliary-Classifier GAN for Malware Analysis}

\author{Rakesh Nagaraju\footnotemark[1]\ \ \ 
Mark Stamp\footnotemark[1]\,\,\footnotemark[2]}

\begin{document}

\symbolfootnotetext[1]{Department of Computer Science, San Jose State University}
\symbolfootnotetext[2]{mark.stamp$@$sjsu.edu}

\maketitle

\abstract
Generative adversarial networks (GAN) are a class of powerful 
machine learning techniques,
where both a generative and discriminative model are trained simultaneously. 
GANs have been used, for example, to successfully generate ``deep fake'' images.
A recent trend in malware research consists of treating executables as images 
and employing image-based analysis techniques. In this research, we 
generate fake malware images using  
auxiliary classifier GANs (AC-GAN), and we consider the effectiveness
of various techniques for classifying the resulting images. 
Our results indicate that the resulting multiclass classification problem 
is challenging, yet we can obtain strong results when restricting
the problem to distinguishing between real and
fake samples. While the AC-GAN generated images often
appear to be very similar to real malware images, we
conclude that from a deep learning perspective,
the AC-GAN generated samples do not rise to the level of 
deep fake malware images.

\section{Introduction}

Malware is malicious software that is intentionally designed to do harm. 
The potential dangers of malware include access to private data, which in turn can lead to 
confidential or financial data theft, identity theft, ransomware,
and other problems. 
Those affected by malware attacks can range from large corporations
and government organizations to a typical individual computer user. 
According to McAfee Labs, ``419 malware threats were encountered 
per minute in the second quarter of~2020, an increase of almost~12\% over the previous quarter''~\cite{mcafee}. 
Malware plays a major role in computer crime and information warfare, 
and hence malware research plays a prominent---if not dominant---role in the field 
of cybersecurity.


A recent trend in malware research consists of treating executables as images, which opens the
door to the use of image-based analysis techniques. For example, a malware detector that uses
image features known as ``gist descriptors'' is considered in~\cite{one}. 
Other image-based approaches that have
been used with success in the malware domain include convolution neural networks (CNN) and
extreme learning machines (ELM); see~\cite{ten} and~\cite{thirteen}, respectively.

A generative adversarial network (GAN) is a powerful machine learning concept where both a
generative and discriminative networks are trained simultaneously~\cite{one}. 
GANs have previously been studied in the context of malware images. 
For example, in~\cite{seven} a transfer learning-based GAN method is used to classify 
previously unknown malware---so-called zero-day
malware. In this approach, GANs are used to generate fake malware images that serve
to augment the training data, thereby reducing the required number of training samples.

In this research, we focus on generating realistic fake malware images using GANs, 
and we consider classification of the resulting fake and real images. 
Specifically, we use auxiliary classifier GAN
(AC-GAN), which enables us to work with multiclass data. We first convert malware executables
from a large and diverse malware datasets into images. We train AC-GAN models on these images,
which enables us to generate fake malware images corresponding to each family. To determine the
quality of these fake samples, we train various models, including CNNs and ELMs, to distinguish
between the real and fake samples. The performance of these models provide an indication of the
quality of our fake malware images---the worse the models perform, the better,
in some sense, are our fake malware images.
We also consider the quality of the discriminative models
trained using AC-GANs. In all cases, we experiment with various combinations of real and fake
malware images. 

The remainder of this paper is organized as follows. 
Section~\ref{chap:2} covers relevant related work. 
In Section~\ref{chap:3}, we outline the methodologies used in this project. 
Section~\ref{chap:4} provides details on the datasets and our specific implementation.
Our experimental results appear in Section~\ref{chap:5}, 
while in Section~\ref{chap:6}, we conclude the paper 
and provide a brief discussion of possible avenues for future work.

\section{Related Work}\label{chap:2}

In this section, we selectively survey some of the previous work 
related to malware classification using machine learning techniques. 
The limitations and advantages of various approaches are considered.  

Most malware detectors are based on some form of pattern matching. 
An inherent weakness of such techniques is that
a malware writer can evade detection by altering the underlying pattern. 
Even statistical and machine learning-based malware detectors can be susceptible to 
a wide variety of code obfuscation techniques~\cite{one}. 
Hence, the challenge is to find an efficient approach that provides strong results along with 
robustness, even under such attack scenarios.

In~\cite{nine} deep learning techniques are considered for malware classification.
The results from two different experiments show that deep learning techniques 
achieves better accuracy than standard malware detectors. 
However, these models are costly, particularly in terms of training. 

A semi-supervised malware detection approach is proposed~\cite{santos2011semi}. 
Here, the authors use a technique that they refer to 
as ``learning with local and global consistency'' 
to reduce dependency on labeled data. 
In~\cite{cakir2018malware}, another popular deep learning model, Word2Vec, 
is used for malware representation. Paired with a gradient search algorithm,
this method achieves an accuracy of about~94\%. 
However, for both this model, the training time is high.

In~\cite{seven}, the authors
show that the generative aspect of GANs can be used to improve malware classification. 
The article~\cite{three} proposes a GAN-based model, 
denoted as MalGAN, that generates fake malware, 
which the authors claim are undetectable by state-of-the-art techniques. 
In~\cite{four}, MalGAN is extended to ``improved MalGAN,''
which additionally learns benign features. These approaches 
were trained on a variety of features, including opcodes. 
Experiments in~\cite{six} show that a deep convolution GAN 
can enable training with limited data,
while in~\cite{seven}, deep learning GAN models are used 
to produce images that appear to be malware samples
visualized as images~\cite{dcgan}.

In~\cite{eleven}, a conditional GAN is used to produce results
comparable to previous research, while additionally providing more 
control over the image generation. One problem in this case, 
is that the discriminator model cannot be used to classify 
the sample labels, as the labels are passed as a parameter to the model. 

In~\cite{three, four}, malware detection models are trained on a variety
of features, including opcodes.  
Specifically, in~\cite{three}, detectors based on neural networks 
are generated by considering malware features such as opcodes. 
It should be noted that the extraction and processing of opcodes 
is a relatively costly process.  

A recent trend in malware research consists of treating executables as images, 
which opens the door to the use of image-based analysis techniques. 
In~\cite{nataraj2011malware}, the authors develop a procedure 
to convert executable binary files into grayscale images.
In~\cite{conti2010visual}, the authors determine the parts of an executable 
(\texttt{.text}, \texttt{.data},. etc.) 
based on image structure. As mentioned above,
a malware detector that relies on image features 
known as gist descriptors is described in~\cite{one},
where experiments show that using malware images
results in a relatively robust detection technique. 

Deep learning techniques including recurrent neural networks (RNN) and 
convolutional neural networks (CNN) are applied to malware images
in~\cite{sun2018deep}. Good accuracies are observed for these approaches, 
which further supports the use of images for malware analysis.
Other image-based malware research involves CNNs and 
extreme learning machines (ELM);
see~\cite{ten} and~\cite{thirteen}, respectively. 

The literature to date clearly shows that deep learning models 
applied to malware images can yield strong results.
In this vein, we build on previous GAN-based malware research.

\section{Methodology}\label{chap:3}

The goal of this research is to create realistic-looking fake malware images,
and then analyze these images using various learning techniques.
We achieve this using GANs, in particular, AC-GANs. 
The real malware images are fed through AC-GAN which, as part of
its training, learns to generate fake malware images (generator)
as well as to discriminate between real and fake (discriminator).  
Once, we generate these fake malware images, we analyze their 
quality by various means. 

\subsection{Data}\label{sec:3.1} 

We use two distinct datasets in this research. First,
the MalImg dataset contains more than~9000 malware images 
belonging to~25 distinct families~\cite{nataraj2011malware}. The MalImg
dataset has been widely studied in image-based malware research.
We have also constructed a new malware image dataset that we refer to 
as MalExe. The MalExe dataset
is derived from more than~24,000 executables belonging to~18 families---we 
obtained the executables from~\cite{dang2021malware}. 

The malware families in the MalImg and MalExe datasets
are listed in Tables~\ref{tab:dataset_info} and~\ref{tab:dataset_info2}, respectively.
Since the MalExe files are executable binaries, we convert them 
to images using a similar approach as in~\cite{ten,nataraj2011malware}.
We discuss this conversion process in more detail below.

\begin{table}[htb]
	\caption{Details of MalExe dataset}\label{tab:dataset_info}
	\centering
        \adjustbox{scale=0.8}{
	\begin{tabular}{c|c|c}\midrule\midrule
			Family & Type & Description\\ \toprule
			\texttt{Alureon} & Trojan & Provides access to confidential data~\cite{alureon}\\
			\texttt{BHO} & Trojan & Performs malicious activities~\cite{bho}\\
			\texttt{CeeInject} & VirTool & Obfuscated code performs any actions~\cite{ceeinject}\\
			\texttt{Cycbot} & Backdoor & Provides control of a system to a server~\cite{cycbot}\\
			\texttt{DelfInject} & VirTool & Provides access to sensitive information~\cite{delfinject}\\
			\texttt{FakeRean} & Rogue & Raises false vulnerabilities~\cite{fakerean}\\
			\texttt{Hotbar} & Adware & Displays ads on browsers~\cite{hotbar}\\
			\texttt{Lolyda.BF} &  Password Stealer & Monitors and sends user's 
				network activity~\cite{lolyda}\\
			\texttt{Obfuscator} & VirTool & Obfuscated code, hard to detect~\cite{obfuscator}\\
			\texttt{OnLineGames} & Password Stealer & Acquires login information of 
				online games~\cite{onlinegames}\\
			\texttt{Rbot} & Backdoor & Provides control of a system~\cite{rbot}\\
			\texttt{Renos} & Trojan Downloader & Raises false warnings~\cite{renos}\\
			\texttt{Startpage} & Trojan & Change browser 
				homepage/other malicious actions~\cite{startpage}\\
			\texttt{Vobfus} & Worm & Download malware and spreads it 
				through USB~\cite{vobfus}\\
			\texttt{Vundo} & Trojan Downloader & Downloads malware using 
				pop-up ads~\cite{vundo}\\ 
			\texttt{Winwebsec} & Rogue & Raises false 
				vulnerabilities~\cite{winwebsec}\\
			\texttt{Zbot} & Password Stealer & Steals personal 
				information through spam emails~\cite{zbot}\\
			\texttt{Zeroaccess} &Trojan Horse & Downloads malware on 
				host machines~\cite{zeroaccess}\\ 
			\midrule\midrule
	\end{tabular}
	}
\end{table}

\begin{table}[htb]
	\caption{Details of MalImg dataset}\label{tab:dataset_info2}
	\centering
        \adjustbox{scale=0.8}{
	\begin{tabular}{c|c|c}\midrule\midrule
			Family & Type & Description\\ \toprule
			\texttt{Adialer.C} & Dialer & Perform malicious activities~\cite{adialer}.\\
			\texttt{Agent.FYI} & Backdoor & Exploits DNS server service~\cite{agentfyi}.\\
			\texttt{Allaple.A} & Worm & Performs DoS attacks~\cite{allaplea}.\\
			\texttt{Allaple.L} & Worm & Worm that spreads itself~\cite{allaplel}.\\
			\texttt{Alureon.gen!J} & Trojan & Modifies DNS settings~\cite{alureongenj}.\\
			\texttt{Autorun.K} & Worm:AutoIT & Worm that spreads itself~\cite{autorunk}.\\
			\texttt{C2LOP.gen!g} & Trojan & Changes browser settings~\cite{c2lopgeng}.\\
			\texttt{C2LOP.P} & Trojan & Modifies bookmarks, popup adds~\cite{c2lopp}.\\
			\texttt{Dialplatform.B} & Dialer & Automatically dials high 
				premium numbers~\cite{diaplatformb}.\\
			\texttt{Dontovo.A} & Trojan downloader & Download and execute 
				arbitrary files~\cite{dontovoa}.\\
			\texttt{Fakerean} & Rogue & Pretends to scan, but steals data~\cite{fakerean}.\\
			\texttt{Instantaccess} & Dialer & Drops trojan to system~\cite{instantaccess}.\\
			\texttt{Lolyda.AA1} & PWS & Steals sensitive information~\cite{lolydaaa1}.\\
			\texttt{Lolyda.AA2} & PWS & Steals sensitive information~\cite{lolydaaa1}.\\
			\texttt{Lolyda.AA3} & PWS & Steals sensitive information~\cite{lolydaaa1}.\\
			\texttt{Lolyda.AT} & PWS & Steals sensitive information~\cite{lolydaaat}.\\
			\texttt{Malex.gen!J} & Trojan & Allows hacker to perform 
				desired actions~\cite{malesgenj}.\\
			\texttt{Obfuscator.AD} & Trojan Downloader & Allows hacker to 
				perform desired actions~\cite{obfuscator}.\\ 
			\texttt{Rbot!gen} & Backdoor & Allows hacker to perform desired 
				actions~\cite{rbotgen}.\\
			\texttt{Skintrim.N} & Trojan & Allows hacker to perform desired 
				actions~\cite{skintrimn}.\\
			\texttt{Swizzor.gen!E} &Trojan downloader & Downloads and installs 
				unwanted software~\cite{swizzorgene}.\\
			\texttt{Swizzor.gen!I} &Trojan downloader & Downloads and installs 
				unwanted software~\cite{swizzorgeni}.\\
			\texttt{VB.AT} & Worm & Spreads automatically across 
				machines~\cite{vbat}.\\
			\texttt{Wintrim.BX} &Trojan downloader & Download and 
				install other software~\cite{wintrimbx}.\\
			\texttt{Yuner.A} &Worm & Spreads automatically across 
				machines~\cite{yunera}.\\ \midrule\midrule
	\end{tabular}
	}
\end{table}

\begin{figure}[!htb]
\centering
\begin{tabular}{ccc}
	\includegraphics[width=0.25\textwidth]{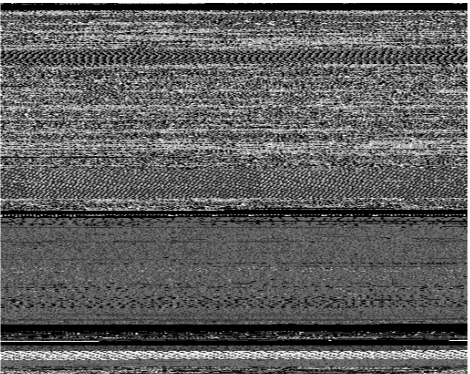}
	& & 
	\includegraphics[width=0.25\textwidth]{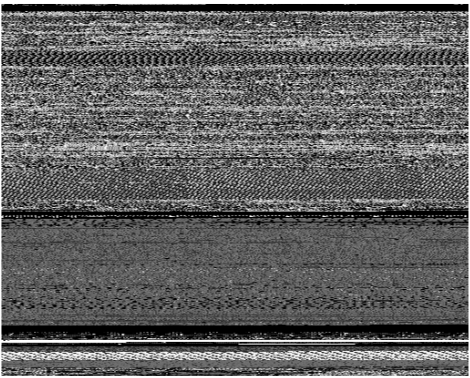}
	\\
	\multicolumn{3}{c}{(a) Examples of \texttt{Adialer.C} from MalImg}
	\\
	\\
	\includegraphics[width=0.225\textwidth]{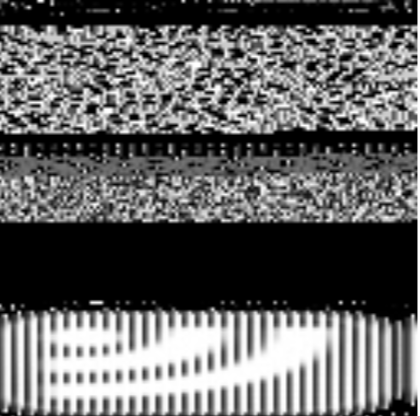}
	& & 
	\includegraphics[width=0.225\textwidth]{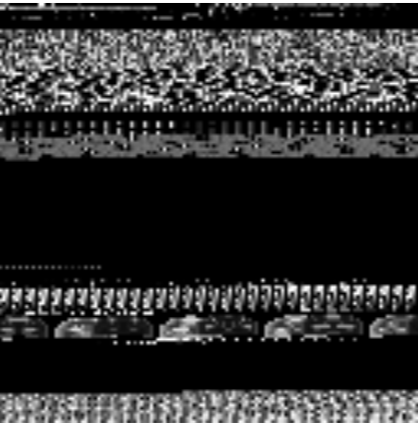}
	\\
	\multicolumn{3}{c}{(b) Examples of \texttt{Obfuscator} from MalExe}
\end{tabular}
\caption{Images from the MalImg and MalExe datasets}\label{fig:401}
\end{figure}

Figure~\ref{fig:401} shows samples of images from the
\texttt{Adialer.C} family of the MalImg dataset and images of the
\texttt{Obfuscator} family from the MalExe dataset. 
For these samples we observe a strong similarity of images within the same family,
and obvious differences in images between different families. This is typical, 
and indicates that image-based analysis should be useful in the malware field.

\subsection{AC-GAN}\label{sec:3.2} 

A generative adversarial network (GAN) is a type of neural network 
that---among many other uses---can generate so-called ``deep fake'' images~\cite{eleven}.
A GAN includes a generator model and a discriminative model
that compete with each other in a min-max game.
Intuitively, this competition will should make both models stronger 
than if each was trained separately, using only the available training data. 
The GAN generator generates fake training samples, with the goal of defeating 
the GAN discriminative model, while the discriminative model tries to distinguish 
real training samples from fake. 

\begin{figure}[!htb]
	\centering
	\includegraphics[width=0.3\textwidth]{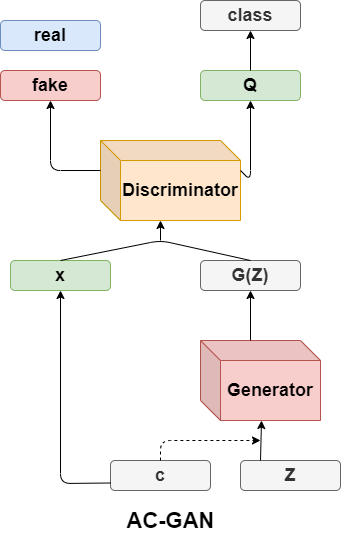}
	\caption{Schematic representation of AC-GAN} 
	\label{fig:402}
\end{figure}

However, a standard GAN is not designed to work with multiclass data. 
Since we have multiclass data, we use
auxiliary-classifier GAN (AC-GAN), which is an enhanced type of GAN
that includes a class label in the generative model.
Additionally, the discriminator predicts both the class label 
and the validity (i.e., real or fake) of a given sample. 
A schematic representation of AC-GAN is given in Figure~\ref{fig:402}. 

For the research in this paper, the key aspect of AC-GAN is that
it enables us to have control of the class of any image that we generate.
We will also make use of AC-GAN discriminative models, as they will serve 
as a baseline for comparison to other deep learning techniques---specifically,
CNNs and ELMs.

\subsection{Evaluation Plan}\label{sec:3.3} 

Once, we have trained and tested our AC-GAN model, we need to evaluate 
the quality of the fake images.
To do this, we compare the AC-GAN classifier to CNN and ELM
models trained on real and fake samples. The remainder of
this section is devoted to a brief introduction to CNNs and ELMs. 

\subsubsection{CNN} 

A convolutional neural network (CNN) is loosely based on the way
that a human perceives an image. 
We first recognize edges, the general shape, texture, and so on,
eventually building up to the point where we can identify a complex object.

\begin{figure}[!htb]
	\centering
	\includegraphics[width=0.45\textwidth]{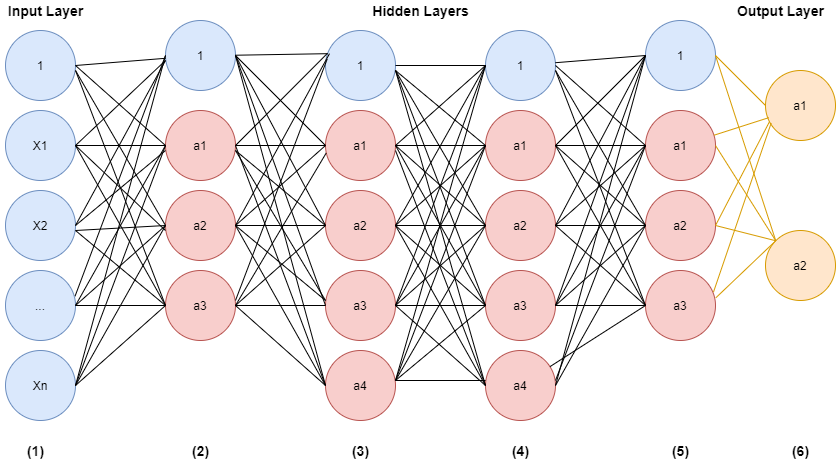}
	\caption{A generic CNN} 
	\label{fig:403}
\end{figure}

A CNN is a feed-forward neural network that includes convolution layers
in which convolutions (i.e., filters) are applied to produce higher level
feature maps. CNNs typically also include pooling layers that primarily serve to
reduce the dimensionality of the problem via downsampling. 
CNNs also typically have a final
fully-connected layer, where all inputs from previous layers are mapped to 
all possible outputs. A generic CNN architecture
is given in Figure~\ref{fig:403}.

For our experiments, we will use the specific CNN architecture and hyperparameters 
specified in~\cite{ten}. The CNN experiments performed in our research involve
malware images, and the specific architecture that we adopt was optimized for precisely
this problem.

\subsubsection{ELM} 

\begin{figure}[ht!]
	\centering
	\includegraphics[width=0.375\textwidth]{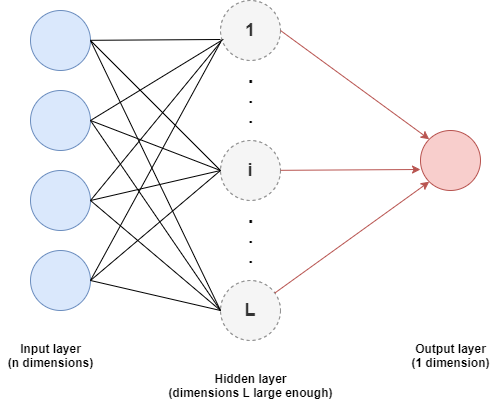}
	\caption{Schematic representation of ELM} 
	\label{fig:404}
\end{figure}

A so-called extreme learning machine (ELM) is a feedforward
deep learning architecture that does not require any 
back-propagation. The weights and biases in the hidden layers
of an ELM are assigned at random, and only the output 
weights are determined via training. Due to this simple structure,
an ELM can be trained using a straightforward equation solving technique---specifically,
the Moore-Penrose generalized inverse.
Thus, ELMs are extremely efficient to train.
A schematic representation of a generic ELM can be seen in
Figure~\ref{fig:404}.

For our experiments, we will use ELM models with parameters as specified 
in~\cite{ten}. As with the CNN experiments mentioned above,
the experiments performed in our research involve
malware images, and the specific ELM architecture that we use was optimized for
this specific problem.

To evaluate the quality of our AC-GAN generated images, 
we first divide the real and fake images into training and testing sets.
Then we train a CNN (respectively, ELM) on the training dataset.
Once, the CNN (respectively, ELM) has been trained, we predict class labels
and determine the accuracy of the predictions. 
The worse the classification accuracy of the CNN (respectively, ELM), the 
better are our AC-GAN generated fake images. We also want to
compare the accuracy of the CNN  and ELM models to the AC-GAN
discriminator. Note that we consider 
each real family and each fake family as a separate class, in effect
doubling the number of classes from the original dataset.

\section{Implementation}\label{chap:4}

In this section, we present details on the implementation
of the techniques discussed in Section~\ref{chap:3}. 
All of our learning techniques have been implemented in
Python using PyTorch and Keras, with the experiments 
run on Google Colab Pro under a local Windows OS.
The precise specifications are given in the Table~\ref{tab:specs}.


\begin{table}[!htb]
	\caption{Environment specifications}\label{tab:specs}
	\centering
	\adjustbox{scale=0.85}{
	\begin{tabular}{c|c}\midrule\midrule
			Specification & Description\\ \toprule
			\multirow{4}{*}{Local machine} & 
			Windows OS \\ 
			& Intel(R) Core(TM) i7-9750H CPU @~2.60GHz \\ 
			& 16.0 GB RAM\\ 
			& NVIDIA GeForce RTX~2060 14GB GPU\\ \midrule
			\multirow{3}{*}{Google Colab Pro} &
			24 hours available runtime \\ 
			& 25 GB memory\\ 
			& T4 and P100 GPUs\\  \midrule
			\multirow{5}{*}{Software} &
			PyTorch \\ 
			& Keras \\ 
			& Numpy \\ 
			& Scipy \\ 
			& PIL \\ 
			\midrule \midrule
	\end{tabular}
	}
\end{table} 

In the remainder of this section 
we provide details on the pre-processing applied to the datasets
used in our experiments, we outline our AC-GAN training process,
and we discuss the training and testing of our CNN and ELM evaluation models.
Then in Section~\ref{chap:5} we present out experimental results.

\subsection{Dataset Analysis and Conversion}\label{sec:4.1}

As mentioned above, we experiment with two distinct datasets. 
In both cases, we use the \texttt{ImageDataGenerator} and \texttt{Dataloader} 
modules from Keras (in PyTorch) to extract images and labels from the data.
Additionally we use the \texttt{transforms} functions to compose our 
pre-processing requirement. 

\subsubsection{Datasets}

The first dataset we consider is the well-known
MalImg dataset, which was originally described in~\cite{eight}. This dataset has
become a standard for comparison in image-based malware research.
The MalImg dataset contains~9339 grayscale images belonging to~25 classes,
where all samples are in the form of images, not executable files.
 
We refer to our second malware image dataset as MalExe, 
and it is of our own creation.
This dataset contains~24,558 malware images belonging to~18 classes.
These samples are in the form of \texttt{exe} files.

Since the MalExe samples are executable binary files, we must
converting them to images. We perform this transform as follows.
i
%
We also construct images by specify a desired size of each 
(square) images as~$n\times n$. We then read the first~$n^2$ bytes 
from a malware binary, and these bytes are viewed 
as~$n\times n$ images of type \texttt{png}.  For example, 
if we specify~$64\times 64$ images, each image is based on the first~4096 bytes
of the corresponding \texttt{exe} file.
In this conversion process, 
we only convert samples that contain a sufficient number of bytes.
In Table~\ref{tab:imagesize}, we see the image counts obtained for the 
MalExe dataset for various image sizes considered. 
Note that for~$512\times 512$ image, we only have~9963 samples 
from~17 classes---the family \texttt{Zeroaccess} has no
samples with at least~$512^2 = 2^{18}$ bytes.

\begin{table}[!htb]
	\caption{MalExe dataset counts}\label{tab:imagesize}
	\centering
	\adjustbox{scale=0.85}{
	\begin{tabular}{c|cc}\midrule\midrule
			Specified image size & Count & Families\\ \toprule
			Standard & 24,652 & 18 \\
			$32\times 32$ & 24,557 & 18 \\
			$64\times 64$ & 24,371 & 18 \\
			$128\times 128$ & 23,369 & 18 \\
			$512\times 512$ & \z\phantom{,}9963 & 17 \\
			\midrule \midrule
	\end{tabular}}
\end{table} 

\begin{figure}[!htb]
	\centering
	\begin{tabular}{ccccc}
	\includegraphics[width=0.15\textwidth]{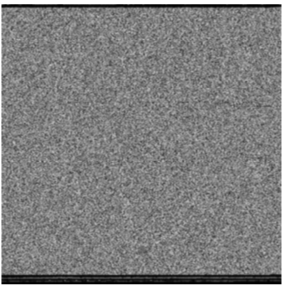}
	&
	\includegraphics[width=0.15\textwidth]{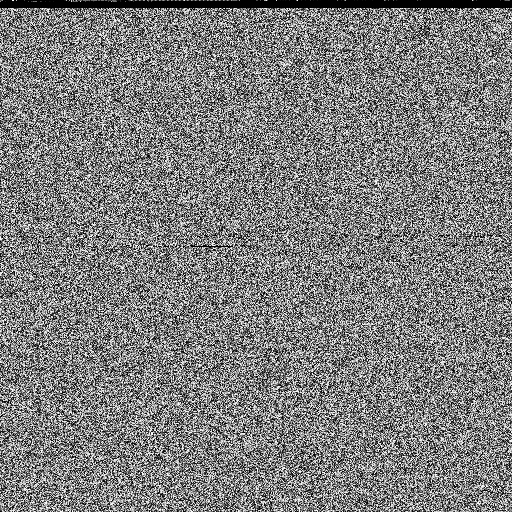}
	&
	\includegraphics[width=0.15\textwidth]{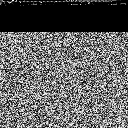}
	&
	\includegraphics[width=0.15\textwidth]{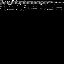}
	&
	\includegraphics[width=0.15\textwidth]{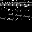}
	\\
	(a) Real
	&
	(b) $512\times 512$
	&
	(c) $128\times 128$
	&
	(d) $64\times 64$
	&
	(e) $32\times 32$
	\end{tabular}
	\caption{Image conversions of an \texttt{Alureon} sample}\label{fig:501_1}
\end{figure}

Figure~\ref{fig:501_1} illustrate images of various sizes for 
one specific sample from the \texttt{Alureon} family. We see that that these different image
construction techniques can provide distinct views of the same data.

\begin{figure}[!htb] 
    \centering
    \begin{tabular}{cc}
	\includegraphics[width=0.4\textwidth]{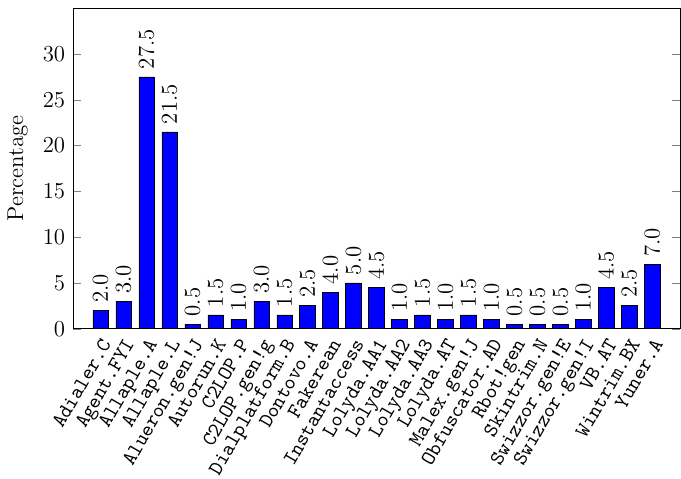}
	&  
	\raisebox{0.725em}{\includegraphics[width=0.4\textwidth]{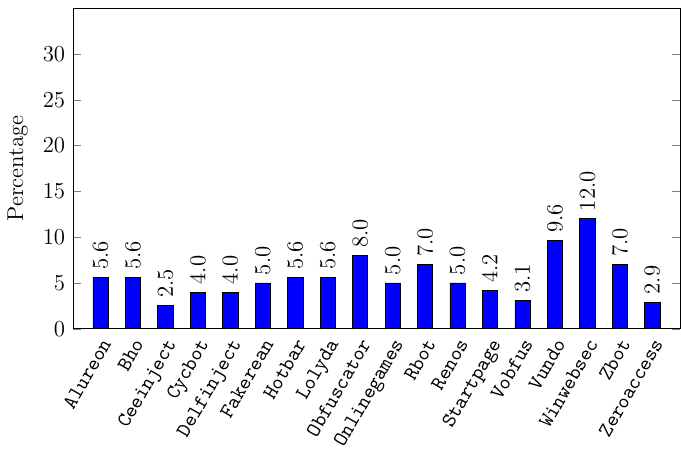}}
	\\
	(a) MalImg
	& 
	(b) MalExe
    \end{tabular}
    \caption{Distribution of samples among families}\label{fig:501}
\end{figure}

In Figure~\ref{fig:501} we give bar graphs showing the distribution of samples
for the MalImg and MalExe datasets. We note that the MalImg dataset is 
highly imbalanced, with the majority of the images belong to \texttt{Allaple.A}, 
\texttt{Allaple.L}, and \texttt{Yuner.A}. 
To deal with this imbalance, we shuffle the data during training and use
balanced accuracy while testing. 

Next, we want to scale the pixel values to the range~$[-1, 1]$ 
in order to match the output of the generator model. This is achieve by
simply calculating the mean pixel value of an entire image
and then subtracting this mean from each pixel and normalizing, 
which gives us a floating point value in the closed interval from~$-1$ to~$+1$
in place of each pixel value.

\subsection{AC-GAN Implementation}\label{sec:4.2}

In this section, we provide additional detail on our implement of AC-GAN.
Recall that our model is generated using Python, PyTorch, and Keras modules.
Also, recall that an AC-GAN includes both a generator and a discriminator. 

\subsubsection{AC-GAN Generator}\label{sub:4.2.1}

Our AC-GAN generator produces a single channel grayscale image 
by plotting random points on a latent space---the latent space 
simply consists of noise drawn from a Gaussian distribution
with~$\mu=0$ and~$\sigma=1$. Additionally, the model 
includes the class label as a parameter. The generator is 
composed as a sequential model. To this sequential model, we 
add a series of deconvolutional layers.
The specific parameters used for the AC-GAN generator 
are given in the Table~\ref{tab:generator_info}.

\begin{table}[!htb]
	\caption{AC-GAN generator construction parameters}\label{tab:generator_info}
	\centering
	\adjustbox{scale=0.85}{
	\begin{tabular}{c|cc}\midrule\midrule
			Layer & Functions & Parameters\\ \toprule
			\multirow{4}{*}{Embedding} & \texttt{Embedding()} 
				& $\mbox{classLabels}\times 100$\\ 
			  & \texttt{Sequential()} &  \\ 
			 & \texttt{Linear()} & in-features: 100; out-features: 131,\\ 
			  & \texttt{Sequential()} &  \\ \midrule
			\multirow{4}{*}{$1^{\stst}$ convolutional} 
				& \texttt{BatchNormal2d()} & in: 128; momentum: 0.1\\ 
			 & \texttt{Upsample()} &Scale factor: 2.0\\ 
			& \texttt{Conv2d()} & in: 128; out: 128; kernel: (3,3); \\  
			&        & stride: (1,1); padding: (1,1)\\  \midrule
			\multirow{5}{*}{$2^{\ndnd}$ convolutional}  
				& \texttt{BatchNormal2d()} & in: 128; momentum: 0.1\\ 
			 & \texttt{LeakyReLU()} & negative\-slope: 0.2\\ 
			 & \texttt{Upsample()} & Scale factor: 2.0\\ 
			 & \texttt{Conv2d()} & in: 128; out: 64; kernel: (3,3); \\ 
			&        & stride: (1,1); padding: (1,1)\\ \midrule
			\multirow{4}{*}{$3^{\rdrd}$ convolutional}  
				& \texttt{BatchNormal2d()} & in: 64; momentum: 0.1\\ 
			 & \texttt{LeakyReLU()} & negative\-slope: 0.2\\ 
			 & \texttt{Conv2d()} & in: 64; out\-channels: 1; kernel: (3,3); \\ 
			&        & stride: (1,1); padding: (1,1)\\ \midrule
			Output & \texttt{Tanh()}  &Scale factor: 2.0\\
			\midrule\midrule
	\end{tabular}
	}
\end{table}

\subsubsection{AC-GAN Discriminator}\label{sub:4.2.2}

The discriminator model discriminates 
between the original and fake images, 
while predicting the class label. 
The generator and discriminator both deal with cross-entropy loss---the 
generator attempts to minimize binary cross-entropy loss, 
while the discriminator tries to maximize this loss.
The discriminator parameters used in our experiments
are given in Table~\ref{tab:discriminator_info}.

\begin{table}[!htb]
	\caption{AC-GAN discriminator construction parameters}\label{tab:discriminator_info}
	\centering
	\adjustbox{scale=0.85}{
	\begin{tabular}{c|cc}\midrule\midrule
			Layer & Functions & Parameters\\ \toprule
			Input & \texttt{Sequential()} &  \\ \midrule
			\multirow{2}{*}{$1^{\stst}$ deconvolutional}  
				& \texttt{Conv2d()} & in: 1; out: 16; kernel: (3,3); \\  
			&        & stride: (2,2); padding: (1,1)\\  \midrule
			\multirow{4}{*}{$2^{\ndnd}$ deconvolutional}  
				& \texttt{LeakyReLU()} & negative\-slope: 0.2\\ 
			& \texttt{Dropout2d()} & rate: 0.25\\ 
			& \texttt{Conv2d()} & in: 16; out: 32; kernel: (3,3); \\  
			&        & stride: (2,2); padding: (1,1)\\ \midrule
			\multirow{5}{*}{$3^{\rdrd}$ deconvolutional}  
				& \texttt{LeakyReLU()} & negative\-slope: 0.2\\ 
			& \texttt{Dropout2d()} & rate: 0.25\\ 
			& \texttt{BatchNormal2d()} & in: 32; momentum: 0.1\\ 
			& \texttt{Conv2d()} & in: 32; out: 64; kernel: (3,3); \\  
			&        & stride: (2,2); padding: (1,1)\\ \midrule
			\multirow{7}{*}{$4^{\thth}$ deconvolutional}  & \texttt{LeakyReLU()} 
				& negative\-slope: 0.2\\ 
			& \texttt{Dropout2d()} & rate: 0.25\\ 
			& \texttt{BatchNormal2d()} & in: 64; momentum: 0.1\\ 
			& \texttt{Conv2d()} & in: 64; out: 128; kernel: (3,3); \\  
			&        & stride: (2,2); padding: (1,1)\\ 
		 	& \texttt{LeakyReLU()} & negative\-slope: 0.2\\ 
			& \texttt{Dropout2d()} & rate: 0.25\\ 
			& \texttt{BatchNormal2d()} & in: 128; momentum: 0.1\\ \midrule
			\multirow{3}{*}{Adversarial} & \texttt{Sequential()} & \\ 
			& \texttt{Linear()} & in-features: 8192; out-features: 1 \\ 
			& \texttt{Sigmoid()} & \\ \midrule
			\multirow{3}{*}{Auxiliary} & \texttt{Sequential()} & \\ 
			& \texttt{Linear()} & in-features: 8192; out-features: 18 \\ 
			& \texttt{Sigmoid()} & \\ 
			\midrule\midrule
	\end{tabular}
	}
\end{table}

Once we have initialized the generator and discriminator models,
the models are then trained. This training process is typical of any
AC-GAN, and hence we omit the details here.
After training, we plot loss graphs to verify training stability. 

\subsection{Evaluation Models}\label{sec:4.3}

To evaluate our AC-GAN generator results, we train CNN and ELM models
on the real and fake images. The better (in some sense)
our AC-GAN generated fake images, the worse the CNN and ELM 
models should perform.

\subsubsection{CNN Implementation}\label{sub:4.3.1}

CNN models include a fully-connected layer, a convolution layer (or layers), 
and a pooling layer (or layers). 
The parameters used in our specific implementation are 
given in Table~\ref{tab:cnn_info}.
The parameters that awe use in our CNN models are as specified in~\cite{thirteen}. 
Note that due to the imbalance in the MalImg dataset,
we use balanced accuracy. 

\begin{table}[!htb]
	\caption{CNN construction parameters}\label{tab:cnn_info}
	\centering
	\adjustbox{scale=0.85}{
	\begin{tabular}{c|cc}\midrule\midrule
			Layer & Functions & Parameters\\ \toprule
			\multirow{3}{*}{$1^{\stst}$ convolutional} & \texttt{Sequential()} &  \\ 
			& \texttt{Conv2d()} & filters: 30; $\mbox{in} = \mbox{image-size}$; 
				$\mbox{out} = 840$; \\  
			&        & kernel: (3,3); activation:  relu\\ \midrule
			$1^{\stst}$ pooling & \texttt{MaxPooling2D()} & size: (2,2)\\ \midrule
			\multirow{2}{*}{$2^{\ndnd}$ convolutional} & \texttt{Conv2d()} 
					& $\mbox{filters} = 15$; $\mbox{in} = 840$; $\mbox{out} = 4065$; \\  
			&        & kernel: (3,3); activation: relu\\ \midrule
			 \multirow{6}{*}{$2^{\ndnd}$ pooling} & \texttt{MaxPooling2D()} & size: (2,2)\\ 
			 & \texttt{Dropout()} & rate: 0.25\\ 
			& \texttt{Flatten()} & \\ 
			 & \texttt{Dense()} & units: 128; out: 376,448; activation: relu\\ 
			 & \texttt{Dropout()} & rate: 0.5\\ \midrule
			\multirow{2}{*}{Other} & \texttt{Dense()} & units: 50; out: 6450; 
				activation: relu\\ 
			 & \texttt{Dense()} & units: num-of-classes; activation: softmax\\ \midrule 
			--- & Loss & categorical cross entropy \\ 
			--- & Optimizer & \texttt{Adam} \\
			\midrule\midrule
	\end{tabular}
	}
\end{table}

\subsubsection{ELM Implementation}\label{sub:4.3.2}

Any ELM includes an initial input layer, a final output layer, and in between these two layers,
there is a hidden layer. The hidden layer weights are assigned at random, with only the output
layer weights determined via training. For an ELM, the only parameter is the
number of hidden units, and we use the value specified in~\cite{ten},
namely~5000.

\section{Experimental Results}\label{chap:5}

Here, we first consider the use of AC-GAN to generate fake malware 
images of various sizes. As part of these experiments, we 
also consider the discriminative ability of AC-GAN discriminator model.

As a followup on our AC-GAN experiments, 
we conduct CNN and ELM experiments in Section~\ref{sec:5.2}.
The purpose of these experiments is to determine how well these deep
learning techniques can distinguish between real malware images
and the AC-GAN generated fake images.

\subsection{AC-GAN Experiments}

We consider AC-GAN experiments to generate fake malware images
of sizes~$32\times 32$, $64\times 64$, and~$128\times 128$.
In each case, we experiment with both the MalImg and MalExe
datasets.

\subsubsection{AC-GAN with $32\times 32$ Images}\label{AC3232}

Our objective here is generate and classify malware images of size~$32\times 32$. 
For the MalImg dataset, which is in the form of images, we resize all of the images 
to~$32\times 32$. We train our AC-GAN model for~1000 epochs with the number of
batches set to~100.
Since there are~9400 MalImg samples in total, we have 94 samples per batch, 
and hence about~94,000 iterations.
Training this model requires about~24 hours on Google Colab Pro. 

In contrast, for the MalExe dataset we read the first~1024 bytes from each binary, 
and treat these bytes as a~$32\times 32$ image. 
We train an AC-GAN model on this dataset
for~500 epochs with the number of batches set to~50.
Since there are~42,266 samples in the MalExe dataset, we
have about~492 samples per batch and requires about~246,000 iterations. 
Training this model also takes about~24 hours on Google Colab Pro. 

\begin{figure}[!htb]
	\centering
	\begin{tabular}{ccc}
	\includegraphics[width=0.4\textwidth]{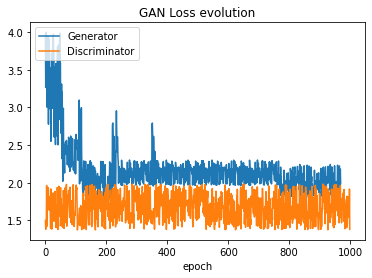}
	& & 
	\includegraphics[width=0.4\textwidth]{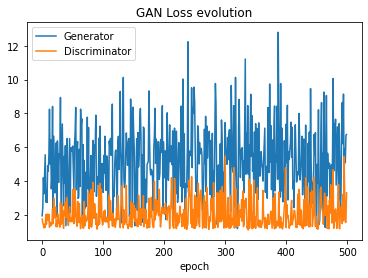}
	\\
	(a) MalImg 
	& &
	(b) MalExe
	\end{tabular}
	\caption{Loss plot for $32\times 32$ images} 
	\label{fig:504}
\end{figure}

Figures~\ref{fig:504}~(a) shows the training loss plots for our AC-GAN generator
and discriminator models when training on the MalImg dataset.
Figures~\ref{fig:504}~(b) shows the corresponding loss plots for the
MalExe dataset. 

From Figure~\ref{fig:504}~(a), we see that both the generator and discriminator stabilizes 
at around epoch~100 for the MalImg experiment. 
The generator spikes up occasionally, but has generally stable loss values, while 
the discriminator loss is more consistent throughout. In contrast, from
Figure~\ref{fig:504}~(b) we see that the MalExe model remains relatively unstable
throughout its~500 iterations.

Our AC-GAN discriminator achieves an accuracy of about~95\%\ in the MalImg
experiment. In contrast, on the MalExe dataset, the AC-GAN
discriminator only attains an accuracy of about~89\%.

\begin{figure}[!htb]
	\centering
	\begin{tabular}{C{3.4cm}C{3.4cm}C{3.4cm}C{3.4cm}}
	\includegraphics[width=0.15\textwidth]{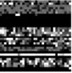}
	&
	\includegraphics[width=0.15\textwidth]{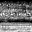}
	&
	\includegraphics[width=0.15\textwidth]{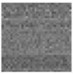}
	&
	\includegraphics[width=0.15\textwidth]{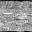}
	\\
	(a) \texttt{\footnotesize C2LOP.P}
	&
	(b) \texttt{\footnotesize C2LOP.P\_fake}
	&
	(c) \texttt{\footnotesize Allaple.L}
	&
	(d) \texttt{\footnotesize Allaple.L\_fake}
	\end{tabular}
	\caption{Real and fake examples from MalImg ($32\times 32$)}\label{fig:506}
\end{figure}

\begin{figure}[!htb]
	\centering
	\begin{tabular}{C{3.4cm}C{3.4cm}C{3.4cm}C{3.4cm}}
	\includegraphics[width=0.15\textwidth]{images/32_Alureon.png}
	&
	\includegraphics[width=0.15\textwidth]{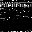}
	&
	\includegraphics[width=0.15\textwidth]{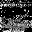}
	&
	\includegraphics[width=0.15\textwidth]{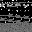}
	\\
	(a) \texttt{\footnotesize Alureon}
	&
	(b) \texttt{\footnotesize Alureon\_fake}
	&
	(c) \texttt{\footnotesize Zeroaccess}
	&
	(d) \texttt{\footnotesize Zeroaccess\_fake}
	\end{tabular}
	\caption{Real and fake examples from MalExe ($32\times 32$)}\label{fig:507}
\end{figure}

Figure~\ref{fig:506} shows a comparison of real and AC-GAN generated 
fake~$32\times 32$ images for 
the families \texttt{C2LOP.P} and \texttt{Allaple.L} from the MalImg dataset. 
Figure~\ref{fig:507} shows a comparison between real and fake images 
for the \texttt{Alureon} and \texttt{Zeroaccess} families from the MalExe data. 
Visually the real and fake images share some characteristics, with the MalExe
fake images being better than the MalImg case. However, the resolution appears
to be too low in all cases. Hence, we perform further AC-GAN experiments 
based on higher resolution images.

\subsubsection{AC-GAN with $64\times 64$ Images}\label{AC6464}

\begin{figure}[!htb]
	\centering
	\begin{tabular}{ccc}
	\includegraphics[width=0.4\textwidth]{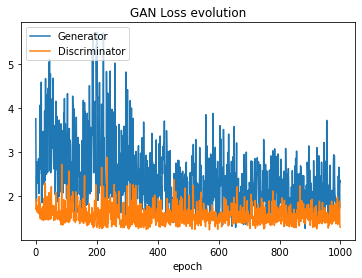}
	& & 
	\includegraphics[width=0.4\textwidth]{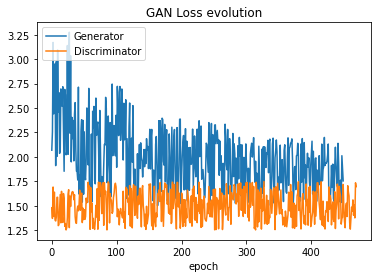}
	\\
	(a) MalImg 
	& &
	(b) MalExe
	\end{tabular}
	\caption{Loss plot for $64\times 64$ images} 
	\label{fig:500}
\end{figure}

Our AC-GAN experiments for~$64\times 64$ images are 
analogous to those for~$32\times 32$
images, as discussed in Section~\ref{AC3232}. Again, the training
time for each dataset is about~24 hours. 
Figures~\ref{fig:500}~(a) and~(b) give the training loss plots for
the MalImg and MalExe experiments, respectively.

From Figure~\ref{fig:500}~(a), we see that the training loss  
stabilizes at around epoch~250 for the MalImg case, while the
MalExe experiment stabilizes at around epoch~100.
In contrast to the~$32\times 32$ case,
the MalExe model becomes reasonably stable after about~125 epochs.

The classification accuracy for the MalImg dataset is about~94\%, while the
AC-GAN achieves a classification accuracy of about~88\%\ on the MalExe
dataset. These results are essentially the same as in the~$32\times 32$ case.

\begin{figure}[!htb]
	\centering
	\begin{tabular}{C{3.4cm}C{3.4cm}C{3.4cm}C{3.4cm}}
	\includegraphics[width=0.15\textwidth]{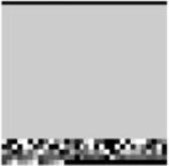}
	&
	\includegraphics[width=0.15\textwidth]{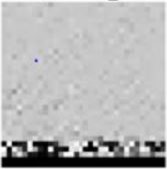}
	&
	\includegraphics[width=0.15\textwidth]{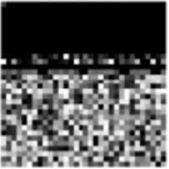}
	&
	\includegraphics[width=0.15\textwidth]{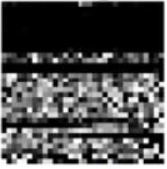}
	\\
	(a) \texttt{\footnotesize Lolyda.AA3}
	&
	(b) \texttt{\footnotesize Lolyda.AA3\_fake} 
	&
	(c) \texttt{\footnotesize Agent.FYI}
	&
	(d) \texttt{\footnotesize Agent.FYI\_fake}
	\end{tabular}
	\caption{Real and fake examples from MalImg ($64\times 64$)}\label{fig:502}
\end{figure}

Again, we compare real and AC-GAN generated fake images. 
Figure~\ref{fig:502} shows the comparison between real and fake images of 
class \texttt{Lolyda.AA3} and \texttt{Agent.FYI} from the MalImg dataset.
We observe that the fake samples in this case are, visually, extremely good.

\begin{figure}[!htb]
	\centering
	\begin{tabular}{C{3.4cm}C{3.4cm}C{3.4cm}C{3.4cm}}
	\includegraphics[width=0.15\textwidth]{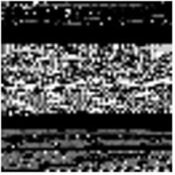}
	&
	\includegraphics[width=0.15\textwidth]{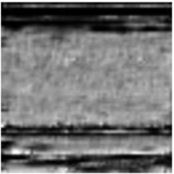}
	&
	\includegraphics[width=0.15\textwidth]{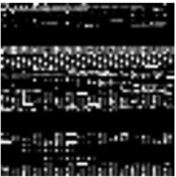}
	&
	\includegraphics[width=0.15\textwidth]{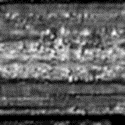}
	\\
	(a) \texttt{\footnotesize Zbot}
	&
	(b) \texttt{\footnotesize Zbot\_fake} 
	&
	(c) \texttt{\footnotesize Vobfus}
	&
	(d) \texttt{\footnotesize Vobfus\_fake}
	\end{tabular}
	\caption{Real and fake examples from MalExe ($64\times 64$)}\label{fig:503}
\end{figure}

In Figure~\ref{fig:503}, we give a comparison between real and fake 
images of class \texttt{Zbot} and \texttt{Vobfus} for the MalExe dataset.
In this case, the MalExe fake samples are surprisingly poor.

\subsubsection{AC-GAN with $128\times 128$ Images}\label{AC128128}

\begin{figure}[!htb]
	\centering
	\begin{tabular}{ccc}
	\includegraphics[width=0.4\textwidth]{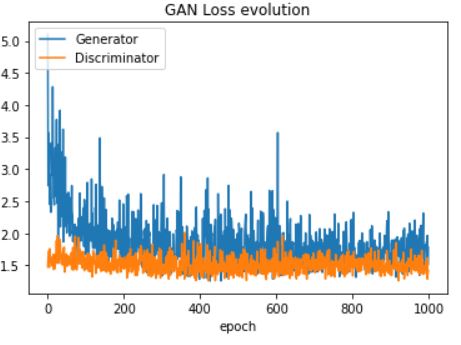}
	& & 
	\includegraphics[width=0.4\textwidth]{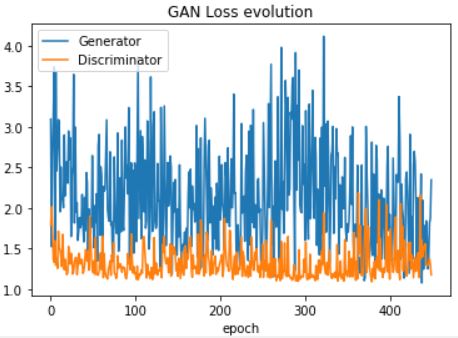}
	\\
	(a) MalImg 
	& &
	(b) MalExe
	\end{tabular}
	\caption{Loss plot for $128\times 128$ images}\label{fig:508}
\end{figure}

We consider AC-GAN experiments based on~$128\times 128$ images. 
These experiments are again analogous to those for the~$32\times 32$
and~$64\times 64$ cases discussed above.
Figures~\ref{fig:508}~(a) and~(b) show the training loss plots 
for AC-GAN trained on the MalImg and MalExe datasets, respectively. 
While the MalImg experiments stabilize, the MalExe experiment
would likely have benefited from additional iterations.

\begin{figure}[!htb]
	\centering
	\begin{tabular}{C{3.4cm}C{3.4cm}C{3.4cm}C{3.4cm}}
	\includegraphics[width=0.15\textwidth]{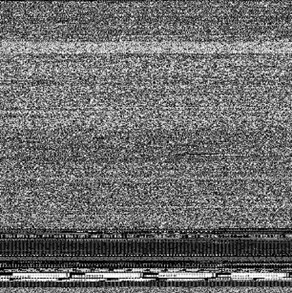}
	&
	\includegraphics[width=0.15\textwidth]{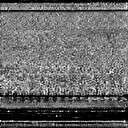}
	&
	\includegraphics[width=0.15\textwidth]{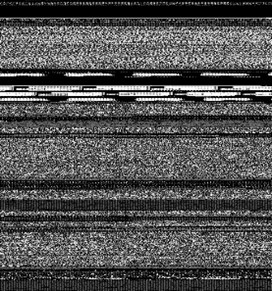}
	&
	\includegraphics[width=0.15\textwidth]{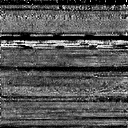}
	\\
	(a) \texttt{\footnotesize Yuner.A}
	&
	(b) \texttt{\footnotesize Yuner.A\_fake} 
	&
	(c) \texttt{\footnotesize VB.AT}
	&
	(d) \texttt{\footnotesize VB.AT\_fake}
	\end{tabular}
	\caption{Real and fake examples from MalImg ($128\times 128$)}\label{fig:511}
\end{figure}

In this case, we attain a maximum classification accuracy from 
the AC-GAN of about~92\%\ for MalImg and about~85\%\ for MalExe.
Figure~\ref{fig:511} shows comparisons of real and fake
\texttt{Yuner.A} and \texttt{VB.AT} from MalImg. 
As in the~$64\times 64$ case, we see that the fake images 
appear to be very good approximations for this dataset.

\begin{figure}[!htb]
	\centering
	\begin{tabular}{C{3.4cm}C{3.4cm}C{3.4cm}C{3.4cm}}
	\includegraphics[width=0.15\textwidth]{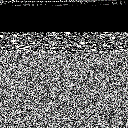}
	&
	\includegraphics[width=0.15\textwidth]{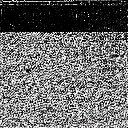}
	&
	\includegraphics[width=0.15\textwidth]{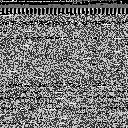}
	&
	\includegraphics[width=0.15\textwidth]{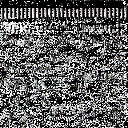}
	\\
	(a) \texttt{\footnotesize Alureon}
	&
	(b) \texttt{\footnotesize Alureon\_fake} 
	&
	(c) \texttt{\footnotesize Zeroaccess}
	&
	(d) \texttt{\footnotesize Zeroaccess\_fake}
	\end{tabular}
	\caption{Real and fake examples from MalExe ($128\times 128$)}\label{fig:510}
\end{figure}

Figure~\ref{fig:510} shows a comparison of real and fake
\texttt{Alureon} and \texttt{Zeroaccess} images from the MalExe data. 
In contrast to the~~$32\times 32$ and~$64\times 64$ cases,
here the fake MalExe images are very good approximations to the real images.

\subsubsection{Summary of AC-GAN Results}\label{sec:ACsum}

Table~\ref{tab:acganaccuracy} gives the discriminative accuracies for each of the AC-GAN 
experiments in Sections~\ref{AC3232} through~\ref{AC128128}.
We see that the results are fairly consistent, irrespective of the
size of the images.

\begin{table}[!htb]
	\caption{AC-GAN discriminator accuracy (rounded to nearest percent)}\label{tab:acganaccuracy}
	\centering
	\adjustbox{scale=0.85}{
	\begin{tabular}{c|cc}\midrule\midrule
			Image Size & Dataset & Accuracy \\ \toprule
			\multirow{2}{*}{$32\times 32$} & MalImg & 95\%\\ 
			    & MalExe & 89\%\\ \midrule
			\multirow{2}{*}{$64\times 64$} & MalImg & 94\%\\ 
			    & MalExe & 88\%\\ \midrule
			\multirow{2}{*}{$128\times 128$} & MalImg & 92\%\\ 
			    & MalExe & 85\%\\ 
			\midrule \midrule
	\end{tabular}
	}
\end{table}

With respect to the visual inspection of the fake images
in Figures~\ref{fig:506} and~\ref{fig:507} (for the~$32\times 32$ case),
Figures~\ref{fig:502} and~\ref{fig:503} (for the~$64\times 64$ case),
and Figures~\ref{fig:511} and~\ref{fig:510} (for the~$128\times 128$ case),
we observed a clear improving trend for larger image sizes.
However, there is a price to be paid for this increased fidelity,
as the training time increases significantly with image size.

\subsection{CNN and ELM Experiments}\label{sec:5.2}

As a first step towards evaluating the quality of the AC-GAN generated images, 
we experiment with CNN and ELM. Specifically, we test the ability of these
two deep learning techniques to distinguish between real malware images
and our AC-GAN generated fake images by treating the real data and 
fake images as distinct classes in multiclass experiments. 
For example, if we consider~10 classes from the MalImg dataset, 
then for our CNN and ELM experiments, we will have~20 classes
consisting of the~10 original families plus another~10 classes
consisting of fake samples from each of the original~10 families. 
In the following sections, 
we separately consider experiments for~$32\times 32$, $64\times 64$, 
and~$128\times 128$ image sizes. 

\subsubsection{CNN and ELM for $32\times 32$ Images}\label{sect:3232}

Here, we consider $32\times 32$ real and fake images and perform
experiments for the MalImg and MalExe datasets. 
For MalExe, we consider all~18 classes and therefore, including classes for the
fake images, we have a total of~36 classes. Our dataset consists of~100 samples for
each class, and hence we have~3600 images. We train our CNN for~3000 
epochs and we generate an ELM with~5000 hidden units. The CNN test accuracy 
is only about~51\%, in spite of a training accuracy of~100\%, which is a sign of overfitting.
The ELM performs slightly worse, achieving an accuracy of~48\%.

Figures~\ref{fig:app32exeCNN} and~\ref{fig:app32exeELM} in 
the appendix give the confusion matrices 
for our CNN and ELM experiments on the MalExe dataset. 
In both cases, we observe that most of the fakes are largely misclassified, 
but this is not the case for all families. 
For example, in the CNN experiments, the fake
\texttt{Vundo} samples are classified correctly
with~100\%\ accuracy, whereas the real
\texttt{Vundo} samples are only classified correctly~33\%\ of the time.

For MalImg, we consider all~25 real classes, which gives~50 classes and
a total of~5000 images. 
Again, our CNN is trained for~3000 epochs and we construct an ELM with~5000 hidden units. 
For the MalImg dataset, our CNN again has a very high training accuracy, 
but achieves a test accuracy of only about~56\%, while our ELM achieves an accuracy 
of about~37\%. The confusion matrices for these experiments
are in Figures~\ref{fig:app32imgCNN} and~\ref{fig:app32imgELM} in the appendix.
Again, we see that the fakes are misclassified at a much higher rate than the real
samples.



\subsubsection{CNN and ELM for $64\times 64$ Images}\label{sect:6464}

In this section, we consider similar experiments as in the previous
section, but based on~$64\times 64$ images. In this case, we
consider~10 of the MalImg families
and the corresponding fake samples, for a total of~20 
classes for each dataset. We again consider~100 images from each class,
and we use~70\%\ of
the samples for training and reserve the remaining~30\%\ for testing.

We train a CNN for~3000 epochs with a batch size of~500 
while for the ELM we use~50,000 hidden units. For the CNN, we
attain~100\%\ training accuracy, but only about~82\%\ test accuracy, which is again
a sign of overfitting. For the ELM, we attain an accuracy of~64\%.
Figures~\ref{fig:app64imgCNN} and~\ref{fig:app64imgELM} in the appendix show the 
confusion matrices for these experiments. From the confusion matrices,
we can see that some images are misclassified as fakes, while some families 
are consistently classified as other families. For both the CNN and ELM, 
we see that most images are misclassified, with the exception of specific families. 
The~$64\times 64$ results---in the form of confusion matrices---for the MalExe dataset
are in Figures~\ref{fig:app64exeCNN} and~\ref{fig:app64exeELM} in the appendix.

\subsubsection{CNN and ELM for $128\times 128$ Images}\label{sect:128128}

In this MalImg experiment, we consider all families in the dataset. In this case, 
we train the CNN for~5000 epochs and generate an ELM with~20,000 hidden units. 
Again, we treat real and fake images as a separate set of classes.
We consider all~18 classes in our MalExe experiments. 

On the MalExe dataset, we achieve~43\%\ test accuracy with the CNN, 
and~52\%\ accuracy with out ELM. Figures~\ref{fig:app128exeCNN} 
and~\ref{fig:app128exeELM} in the appendix 
show the confusion matrices for our CNN and ELM experiments on 
the MalExe data. Similar to other experiments on MalExe, we 
see mostly miscalculation for the CNN. For the ELM, we note that \texttt{Rbot} fake, 
and \texttt{Ceeinject} fake are particularly poor results.
The results of these~$128\times 128$ experiments
again indicate that AC-GAN produces strong fake images.

For the $128\times 128$ MalImg experiments, we consider all classes,
we train the CNN for~3000 epochs, and we generate an ELM with~20,000 hidden units. 
The results for these MalImg experiments are given in 
Figures~\ref{fig:app128imgCNN} and~\ref{fig:app128imgELM} in the appendix.
The CNN achieves only~43\%\ test accuracy, while ELM performs better,
but still only attains an accuracy of~52\%. 

\subsubsection{Discussion of CNN and ELM Experiments}\label{sec:5.3}

\begin{figure}[!htb]
	\centering
	\includegraphics[width=0.75\textwidth]{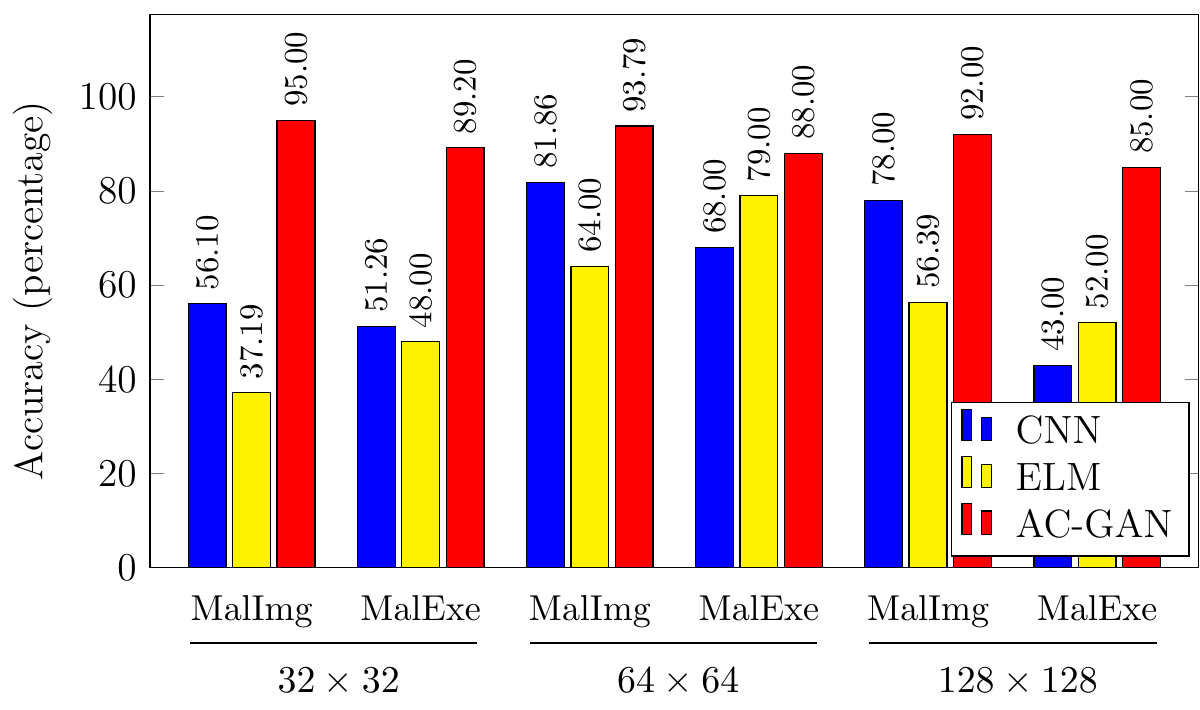}
	\caption{Test accuracy for all experiments}\label{fig:537}
\end{figure}

In Figure~\ref{fig:537} we compare the test accuracies of our CNN and ELM
experiments to our AC-GAN classifier. Here, we observe that the AC-GAN 
models are able to to produce much higher classification rates in all cases.
This shows that while the AC-GAN generator 
is able to produce images that are difficult for
other deep learning techniques to distinguish, the AC-GAN discriminator is
not so easily defeated by these fake images. 
These results suggest that AC-GAN is not only a
source for generating fake malware images, but it is also a powerful
model for discriminating between families---both real and fake.

Finally, we consider the narrower problem of distinguishing real samples from fake samples.
In Figures~\ref{fig:app32condense} through~\ref{fig:app128condense} in the appendix, 
we have ``condensed'' the confusion matrices of
Figures~\ref{fig:app32exeCNN} through~\ref{fig:app128imgELM} 
to better highlight the ability of our CNN and ELM models to distinguish real from fake.
Each of these condensed confusion matrices includes the eight (exhaustive) cases
listed in Table~\ref{tab:ccCases}.

\begin{table}[!htb]
\caption{Condensed confusion matrix cases}\label{tab:ccCases}
\centering
\adjustbox{scale=0.85}{
\begin{tabular}{c|ll}\midrule\midrule
Actual class & Classification & \multicolumn{1}{c}{Description} \\ \toprule
\multirow{4}{*}{\texttt{real}} & \texttt{real-same} & real sample classified correctly \\
			       & \texttt{fake-same} & real sample classified as fake of the same family\\
			       & \texttt{real-other} & real sample classified as a different real family\\
			       & \texttt{fake-other} & real sample classified as a different fake family\\ \midrule
\multirow{4}{*}{\texttt{fake}} & \texttt{real-same} & fake sample classified as real of the same family\\
			       & \texttt{fake-same} & fake sample classified correctly \\
			       & \texttt{real-other} & fake sample classified as a different real family\\
			       & \texttt{fake-other} & fake sample classified as a different fake family\\
			       \midrule\midrule
\end{tabular}
}
\end{table}

\begin{figure}[!htb]
	\centering
	\includegraphics[width=0.75\textwidth]{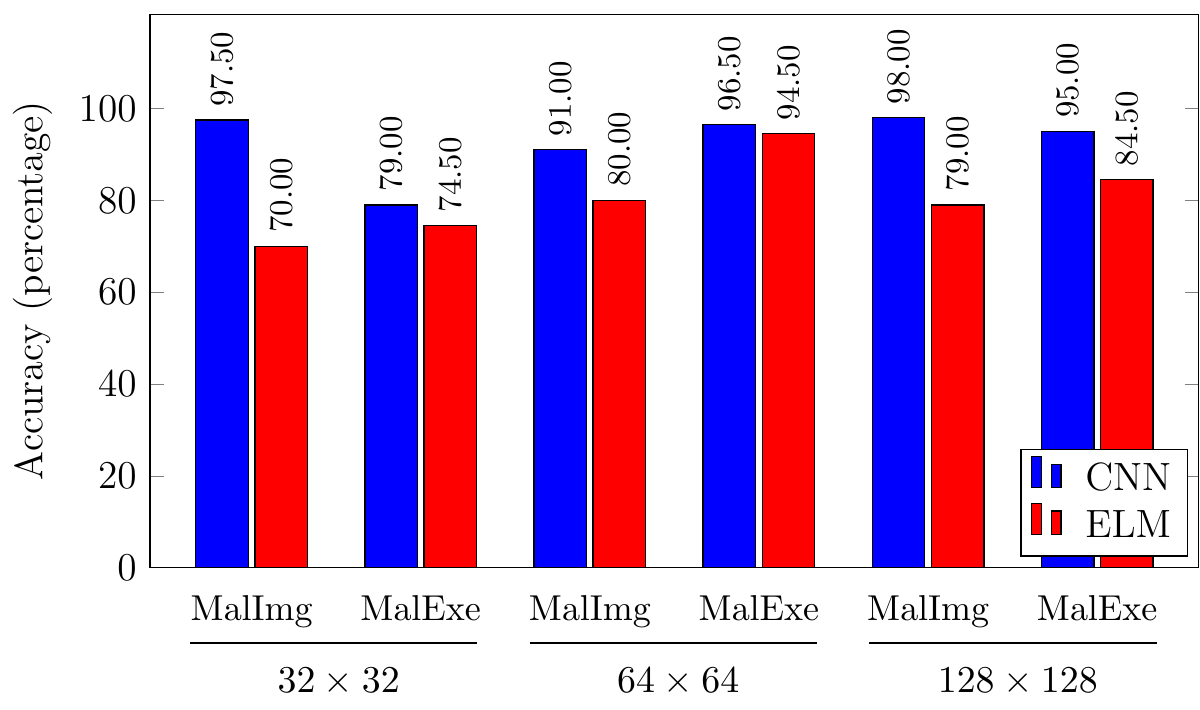}
	\caption{Distinguishing between real and fake}\label{fig:realFake}
\end{figure}

If we are only concerned with the ability of our models to distinguish between real
and fake samples, then any real sample that is classified as 
real---either the correct real family or a different real family---is 
considered a correct classification. Similarly, any fake sample that is
classified as any class of fake is considered a correct classification.
The results in Figure~\ref{fig:realFake} are 
easily obtained from the condensed confusion matrices in 
Figures~\ref{fig:app32condense} through~\ref{fig:app128condense}.
From this perspective, we see that our CNN models always outperform
the corresponding ELM model, and in most cases, the CNN models
perform remarkably well. These results indicate that in spite of the
relatively low accuracies obtained in the multiclass case, most of the
errors are within the real and fake categories, and not between real
and fake samples. In particular, for the CNN models,
real and fake samples from a specific family are rarely 
confused with each other. This provides strong evidence that
the real and fake categories are substantially different
from each other. Perhaps surprisingly, 
these results strongly suggest that AC-GAN
generated fake malware images do not satisfy the requirements
of ``deep fakes,'' at least not from the perspective of 
evaluation by deep learning techniques.

\section{Conclusion and Future Work}\label{chap:6}

In this research we considered AC-GAN in the context of malware research.
We experimented with a standard malware image dataset (MalImg) and
a larger and more balanced malware image dataset of our
own construction (MalExe). We evaluated the images generated by 
our AC-GAN using CNN and ELM models.

We were not able to reliably classify our AC-GAN generated fake 
malware images from genuine malware images 
using either CNNs or ELMs, but the AC-GAN discriminator provided
good accuracy. However, we also found that CNNs can distinguish
between real and AC-GAN generated fake samples with surprisingly high accuracy.

For future work, more experiments aimed at classifying
real and fake malware images
would be useful. Additional state-of-the-art deep learning models,
such as ResNet152 and VGG-19, could be considered~\cite{prajapati2020empirical}.
In addition, the quest for true ``deep fake'' malware images that
cannot be reliably distinguished from real malware images
appears to be a challenging problem.

In addition, it would be interesting to explore adversarial attacks on image-based malware
detectors. For example, tt would be interesting to quantify the effectiveness of 
such attacks. That is, assuming that 
an attacker is able to corrupt the training data,
what is the minimum percentage of the data that must be modified
to achieve a desired level of degradation in the resulting model?

\bibliographystyle{plain}
\bibliography{references.bib}

\begin{thebibliography}{10}

\bibitem{adialer}
Adialer.c.
\newblock
  \url{https://www.microsoft.com/en-us/wdsi/threats/malware-encyclopedia-description?Name=Trojan:Win32/Adialer.C&threatId=-2147460766}.

\bibitem{agentfyi}
Agent.fyi.
\newblock
  \url{https://www.microsoft.com/en-us/wdsi/threats/malware-encyclopedia-description?Name=Exploit:Win32/Siveras.A}.

\bibitem{allaplea}
Allaple.a.
\newblock
  \url{https://www.microsoft.com/en-us/wdsi/threats/malware-encyclopedia-description?Name=worm:win32/allaple.a&ThreatID=2147574777}.

\bibitem{allaplel}
Allaple.l.
\newblock
  \url{https://www.microsoft.com/en-us/wdsi/threats/malware-encyclopedia-description?Name=Worm:Win32/Allaple.L}.

\bibitem{alureon}
Alureon.
\newblock
  \url{https://www.microsoft.com/en-us/wdsi/threats/malware-encyclopedia-description?Name=Win32/Alureon}.

\bibitem{alureongenj}
Alureon.gen!j.
\newblock
  \url{https://www.microsoft.com/en-us/wdsi/threats/malware-encyclopedia-description?Name=Trojan:Win32/Alureon.gen!J}.

\bibitem{autorunk}
Autorun.k.
\newblock
  \url{https://www.microsoft.com/en-us/wdsi/threats/malware-encyclopedia-description?Name=Worm:Win32/Autorun.K&threatId=-2147369124}.

\bibitem{bho}
Bho.
\newblock
  \url{https://www.microsoft.com/en-us/wdsi/threats/malware-encyclopedia-description?Name=Trojan:Win32/BHO&threatId=-2147364778}.

\bibitem{c2lopgeng}
C2lop.gen!g.
\newblock
  \url{https://www.microsoft.com/en-us/wdsi/threats/malware-encyclopedia-description?Name=Trojan:Win32/C2Lop.gen!G&threatId=139219}.

\bibitem{c2lopp}
C2lop.p.
\newblock
  \url{https://www.microsoft.com/en-us/wdsi/threats/malware-encyclopedia-description?Name=Trojan:Win32/C2Lop.P}.

\bibitem{cakir2018malware}
Bugra Cakir and Erdogan Dogdu.
\newblock Malware classification using deep learning methods.
\newblock In {\em Proceedings of the {ACMSE} 2018 Conference}, pages 1--5,
  2018.

\bibitem{ceeinject}
Ceeinject.
\newblock
  \url{https://www.microsoft.com/en-us/wdsi/threats/malware-encyclopedia-description?Name=VirTool%3AWin32%2FCeeInject}.

\bibitem{conti2010visual}
Gregory Conti, Sergey Bratus, Anna Shubina, Andrew Lichtenberg, Roy Ragsdale,
  Robert Perez-Alemany, Benjamin Sangster, and Matthew Supan.
\newblock A visual study of primitive binary fragment types.
\newblock
  \url{https://www.semanticscholar.org/paper/A-Visual-Study-of-Primitive-Binary-Fragment-Types-Conti-Bratus/b406e34d0c203deadfb028f14607bfe88e5763ac},
  2010.

\bibitem{cycbot}
Cycbot.
\newblock
  \url{https://www.microsoft.com/en-us/wdsi/threats/malware-encyclopedia-description?Name=Win32/Cycbot&threatId=}.

\bibitem{dang2021malware}
Dennis Dang, Fabio Di~Troia, and Mark Stamp.
\newblock Malware classification using long short-term memory models.
\newblock \url{https://arxiv.org/abs/2103.02746}, 2021.

\bibitem{delfinject}
Delfinject.
\newblock
  \url{https://www.microsoft.com/en-us/wdsi/threats/malware-encyclopedia-description?Name=PWS:Win32/DelfInject&threatId=-
  2147241365}.

\bibitem{diaplatformb}
Diaplatform.b.
\newblock
  \url{https://www.microsoft.com/en-us/wdsi/threats/malware-encyclopedia-description?Name=Dialer:Win32/DialPlatform.B}.

\bibitem{dontovoa}
Dontovo.a.
\newblock
  \url{https://www.microsoft.com/en-us/wdsi/threats/malware-encyclopedia-description?Name=TrojanDownloader:Win32/Dontovo.A&threatId=-2147342037}.

\bibitem{fakerean}
Fakerean.
\newblock
  \url{https://www.microsoft.com/en-us/wdsi/threats/malware-encyclopedia-description?Name=Win32/FakeRean}.

\bibitem{hotbar}
Hotbar.
\newblock
  \url{https://www.microsoft.com/en-us/wdsi/threats/malware-encyclopedia-description?Name=Adware:Win32/Hotbar&threatId=
  6204}.

\bibitem{three}
Weiwei Hu and Ying Tan.
\newblock Generating adversarial malware examples for black-box attacks based
  on {GAN}.
\newblock \url{https://arxiv.org/abs/1702.05983}, 2017.

\bibitem{dcgan}
Nathan Inkawhich.
\newblock {PyTorch} {DCGAN} tutorial.
\newblock
  \url{https://pytorch.org/tutorials/beginner/dcgan_faces_tutorial.html}.

\bibitem{instantaccess}
Instantaccess.
\newblock
  \url{https://www.microsoft.com/en-us/wdsi/threats/malware-encyclopedia-description?name=dialer:win32/instantaccess}.

\bibitem{ten}
Mugdha Jain.
\newblock Image-based malware classification with convolutional neural networks
  and extreme learning machines.
\newblock \url{https://scholarworks.sjsu.edu/etd\_projects/900/}, 2019.

\bibitem{four}
Masataka Kawai, Kaoru Ota, and Mianxing Dong.
\newblock Improved {MalGAN}: Avoiding malware detector by leaning cleanware
  features.
\newblock In {\em 2019 International Conference on Artificial Intelligence in
  Information and Communication}, ICAIIC, pages 040--045, 2019.

\bibitem{six}
Jin-Young Kim, Seok-Jun Bu, and Sung-Bae Cho.
\newblock Malware detection using deep transferred generative adversarial
  networks.
\newblock In {\em International Conference on Neural Information Processing},
  pages 556--564, 2017.

\bibitem{nine}
David Kornish, Justin Geary, Victor Sansing, Soundararajan Ezekiel, Larry
  Pearlstein, and Laurent Njilla.
\newblock Malware classification using deep convolutional neural networks.
\newblock In {\em 2018 IEEE Applied Imagery Pattern Recognition Workshop},
  AIPR, pages 1--6, 2018.

\bibitem{lolyda}
Lolyda.
\newblock
  \url{https://www.microsoft.com/en-us/wdsi/threats/malware-encyclopedia-description?Name=PWS%3AWin32%2FLolyda.BF}.

\bibitem{lolydaaa1}
Lolyda.aa1.
\newblock
  \url{https://www.microsoft.com/en-us/wdsi/threats/malware-encyclopedia-description?Name=PWS:Win32/Lolyda.AA&threatId=-2147345828}.

\bibitem{lolydaaat}
Lolyda.at.
\newblock
  \url{https://www.microsoft.com/en-us/wdsi/threats/malware-encyclopedia-description?Name=PWS:Win32/Lolyda.AT&ThreatID=2147627867}.

\bibitem{seven}
Yan Lu and Jiang Li.
\newblock Generative adversarial network for improving deep learning based
  malware classification.
\newblock In {\em 2019 Winter Simulation Conference}, WSC, pages 584--593,
  2019.

\bibitem{eight}
Adam Lutz, Victor F.~Sansing III, Waleed~E. Farag, and Soundararajan Ezekiel.
\newblock Malware classification using fusion of neural networks.
\newblock In Misty Blowers, Russell~D. Hall, and Venkateswara~R. Dasari,
  editors, {\em Disruptive Technologies in Information Sciences~{II}}, pages
  165--170. SPIE, 2019.

\bibitem{malesgenj}
Malex.gen!j.
\newblock
  \url{https://www.microsoft.com/en-us/wdsi/threats/malware-encyclopedia-description?Name=Trojan:Win32/Malex.gen!J}.

\bibitem{mcafee}
{McAfee} 2020 2nd quarter report.
\newblock
  \url{https://www.mcafee.com/enterprise/en-us/lp/threats-reports/apr-2021.html}.

\bibitem{nataraj2011malware}
Lakshmanan Nataraj, Sreejith Karthikeyan, Gregoire Jacob, and Bangalore~S
  Manjunath.
\newblock Malware images: Visualization and automatic classification.
\newblock In {\em Proceedings of the 8th International Symposium on
  Visualization for Cyber Security}, pages 1--7, 2011.

\bibitem{obfuscator}
Obfuscator.
\newblock
  \url{https://www.microsoft.com/en-us/wdsi/threats/malware-encyclopedia-description?Name=Win32/Obfuscator&threatId=}.

\bibitem{eleven}
Augustus Odena, Christopher Olah, and Jonathon Shlens.
\newblock Conditional image synthesis with auxiliary classifier {GAN}s.
\newblock \url{https://arxiv.org/abs/1610.09585}, 2017.

\bibitem{onlinegames}
Onlinegames.
\newblock
  \url{https://www.microsoft.com/en-us/wdsi/threats/malware-encyclopedia-description?Name=PWS%3AWin32%2FOnLineGames},
  journal={Onlinegames}.

\bibitem{prajapati2020empirical}
Pratikkumar Prajapati and Mark Stamp.
\newblock An empirical analysis of image-based learning techniques for malware
  classification.
\newblock In {\em Malware Analysis Using Artificial Intelligence and Deep
  Learning}, pages 411--435. Springer, 2020.

\bibitem{rbot}
Rbot.
\newblock
  \url{https://www.microsoft.com/en-us/wdsi/threats/malware-encyclopedia-description?Name=Win32/Rbot&threatId=}.

\bibitem{rbotgen}
Rbot!gen.
\newblock
  \url{https://www.microsoft.com/en-us/wdsi/threats/malware-encyclopedia-description?Name=Backdoor:Win32/Rbot.gen}.

\bibitem{renos}
Renos.
\newblock
  \url{https://www.microsoft.com/en-us/wdsi/threats/malware-encyclopedia-description?Name=TrojanDownloader:Win32/Renos&threatId=16054}.

\bibitem{santos2011semi}
Igor Santos, Javier Nieves, and Pablo~G Bringas.
\newblock Semi-supervised learning for unknown malware detection.
\newblock In {\em International Symposium on Distributed Computing and
  Artificial Intelligence}, pages 415--422, 2011.

\bibitem{skintrimn}
Skintrim.n.
\newblock
  \url{https://www.microsoft.com/en-us/wdsi/threats/malware-encyclopedia-description?Name=Trojan:Win32/Skintrim.N}.

\bibitem{startpage}
Startpage.
\newblock
  \url{https://www.microsoft.com/en-us/wdsi/threats/malware-encyclopedia-description?Name=Trojan:Win32/Startpage&threatId=
  15435}.

\bibitem{sun2018deep}
Guosong Sun and Quan Qian.
\newblock Deep learning and visualization for identifying malware families.
\newblock {\em IEEE Transactions on Dependable and Secure Computing},
  18(1):283--295, 2021.

\bibitem{swizzorgene}
Swizzor.gen!e.
\newblock
  \url{https://www.microsoft.com/en-us/wdsi/threats/malware-encyclopedia-description?Name=TrojanDownloader%253aWin32%252fSwizzor.gen!E&navV3Index=3},
  key={Swizzor.gen!E}.

\bibitem{swizzorgeni}
Swizzor.gen!i.
\newblock
  \url{https://www.microsoft.com/en-us/wdsi/threats/malware-encyclopedia-description?Name=TrojanDownloader:Win32/Swizzor.gen!I}.

\bibitem{vbat}
Vb.at.
\newblock
  \url{https://www.microsoft.com/en-us/wdsi/threats/malware-encyclopedia-description?Name=Worm:Win32/VB.AT}.

\bibitem{vobfus}
Vobfus.
\newblock
  \url{https://www.microsoft.com/en-us/wdsi/threats/malware-encyclopedia-description?Name=Win32/Vobfus&threatId=}.

\bibitem{vundo}
Vundo.
\newblock
  \url{https://www.microsoft.com/en-us/wdsi/threats/malware-encyclopedia-description?Name=Win32/Vundo&threatId=}.

\bibitem{wintrimbx}
Wintrim.bx.
\newblock
  \url{https://www.microsoft.com/en-us/wdsi/threats/malware-encyclopedia-description?Name=TrojanDownloader:Win32/Wintrim.BX}.

\bibitem{winwebsec}
Winwebsec.
\newblock
  \url{https://www.microsoft.com/en-us/wdsi/threats/malware-encyclopedia-description?Name=Win32/Winwebsec}.

\bibitem{one}
Sravani Yajamanam, Vikash Raja~Samuel Selvin, Fabio~Di Troia, and Mark Stamp.
\newblock Deep learning versus gist descriptors for image-based malware
  classification.
\newblock In Paolo Mori, Steven Furnell, and Olivier Camp, editors, {\em
  Proceedings of the 4th International Conference on Information Systems
  Security and Privacy}, ICISSP 2018, pages 553--561, 2018.

\bibitem{thirteen}
Songqing Yue.
\newblock Imbalanced malware images classification: A {CNN} based approach.
\newblock \url{https://arxiv.org/abs/1708.08042}, 2017.

\bibitem{yunera}
Yuner.a.
\newblock
  \url{https://www.microsoft.com/en-us/wdsi/threats/malware-encyclopedia-description?Name=Worm:Win32/Yuner.A&ThreatID=2147600986}.

\bibitem{zbot}
Zbot.
\newblock
  \url{https://www.microsoft.com/en-us/wdsi/threats/malware-encyclopedia-description?Name=PWS:Win32/Zbot&threatId=-2147368817}.

\bibitem{zeroaccess}
Zeroaccess.
\newblock \url{https://www.symantec.com/security-center/writeup/2011-
  071314-0410-99}.

\end{thebibliography}

\appendix
\setcounter{secnumdepth}{0}

\section{Appendix}

In this appendix, we give the confusion matrices for 
each of the~$32\times 32$, $64\times 64$, and~$128\times 128$ 
experiments discussed in Sections~\ref{sect:3232},  \ref{sect:6464}
and~\ref{sect:128128}, respectively.


\begin{figure}[!htb]
	\centering
	\includegraphics[width=0.85\textwidth]{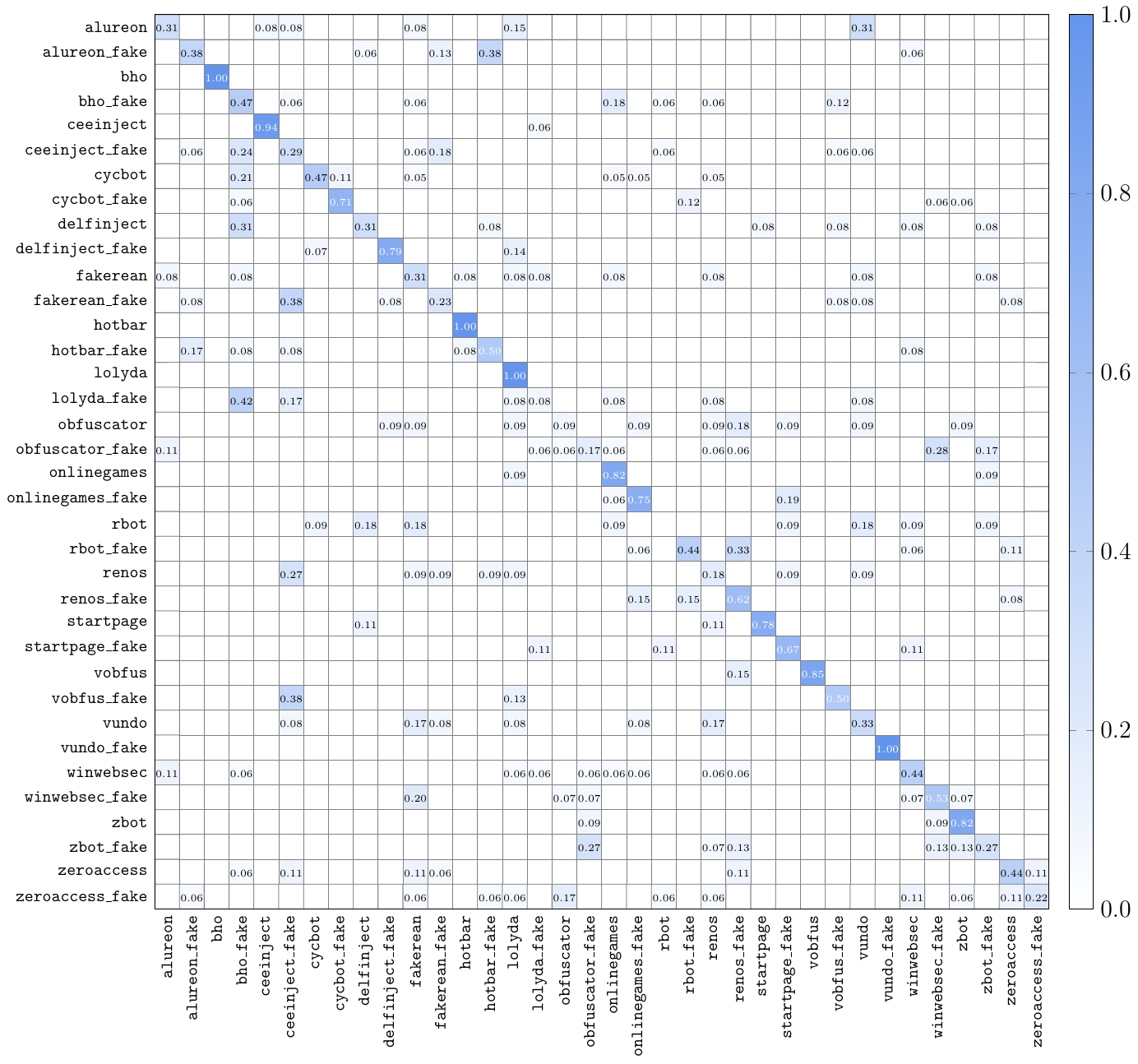}
	\caption{CNN confusion matrix (MalExe $32\times 32$)} 
	\label{fig:app32exeCNN}
\end{figure}

\begin{figure}[!htb]
	\centering
	\includegraphics[width=0.85\textwidth]{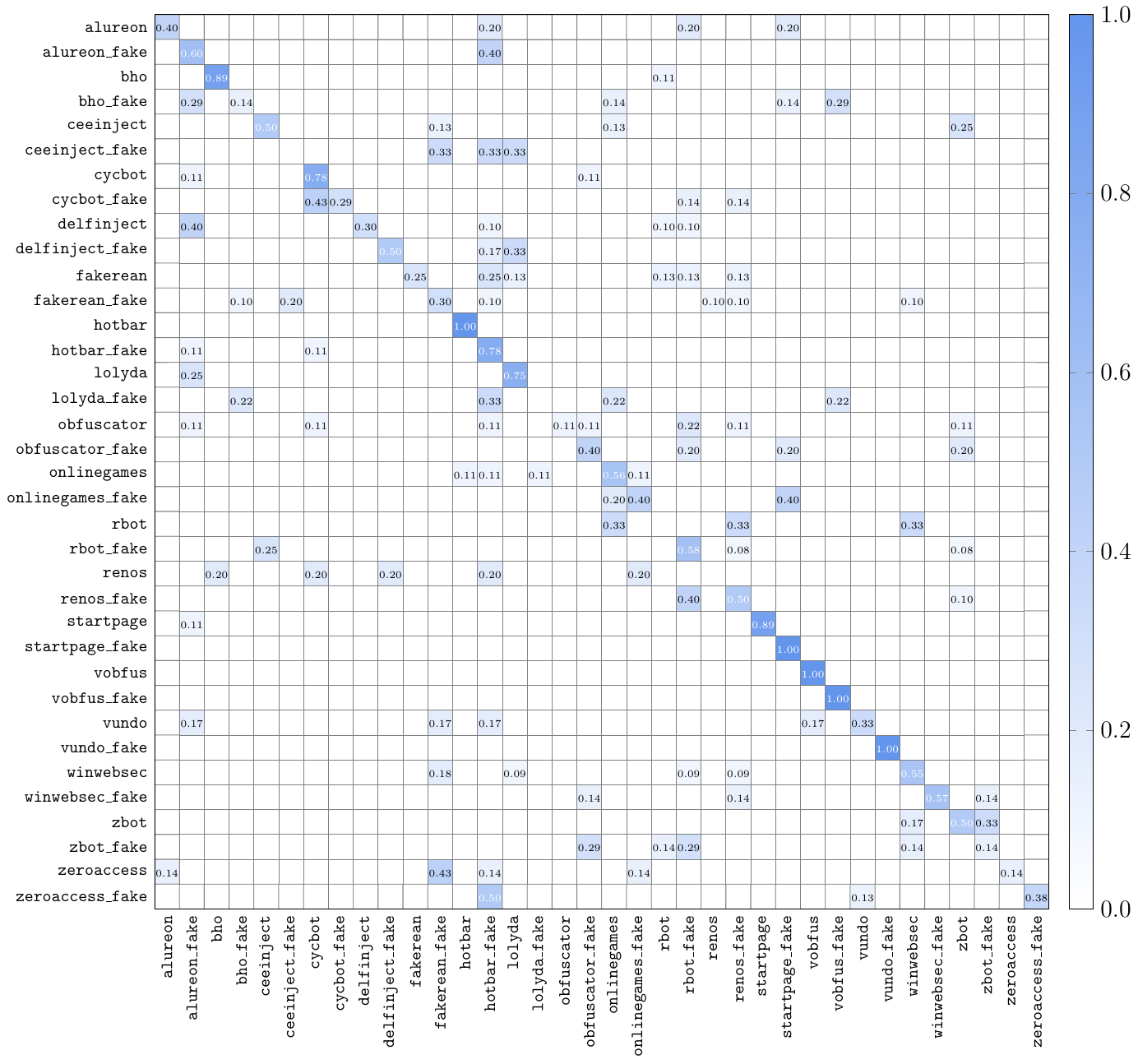}
	\caption{ELM confusion matrix (MalExe $32\times 32$)} 
	\label{fig:app32exeELM}
\end{figure}

\begin{figure}[!htb]
	\centering
	\includegraphics[width=0.85\textwidth]{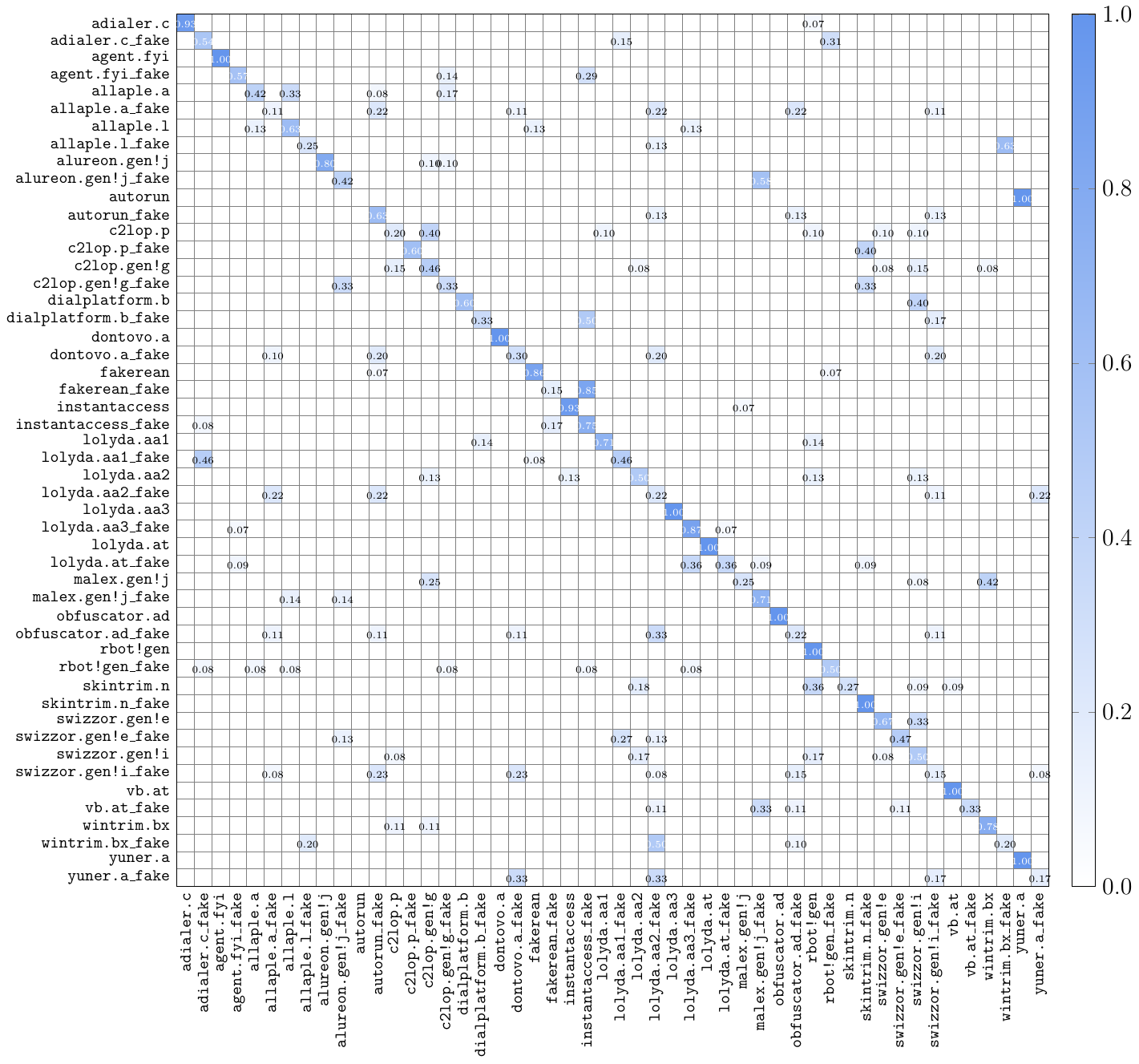}
	\caption{CNN confusion matrix (MalImg $32\times 32$)} 
	\label{fig:app32imgCNN}
\end{figure}

\begin{figure}[!htb]
	\centering
	\includegraphics[width=0.85\textwidth]{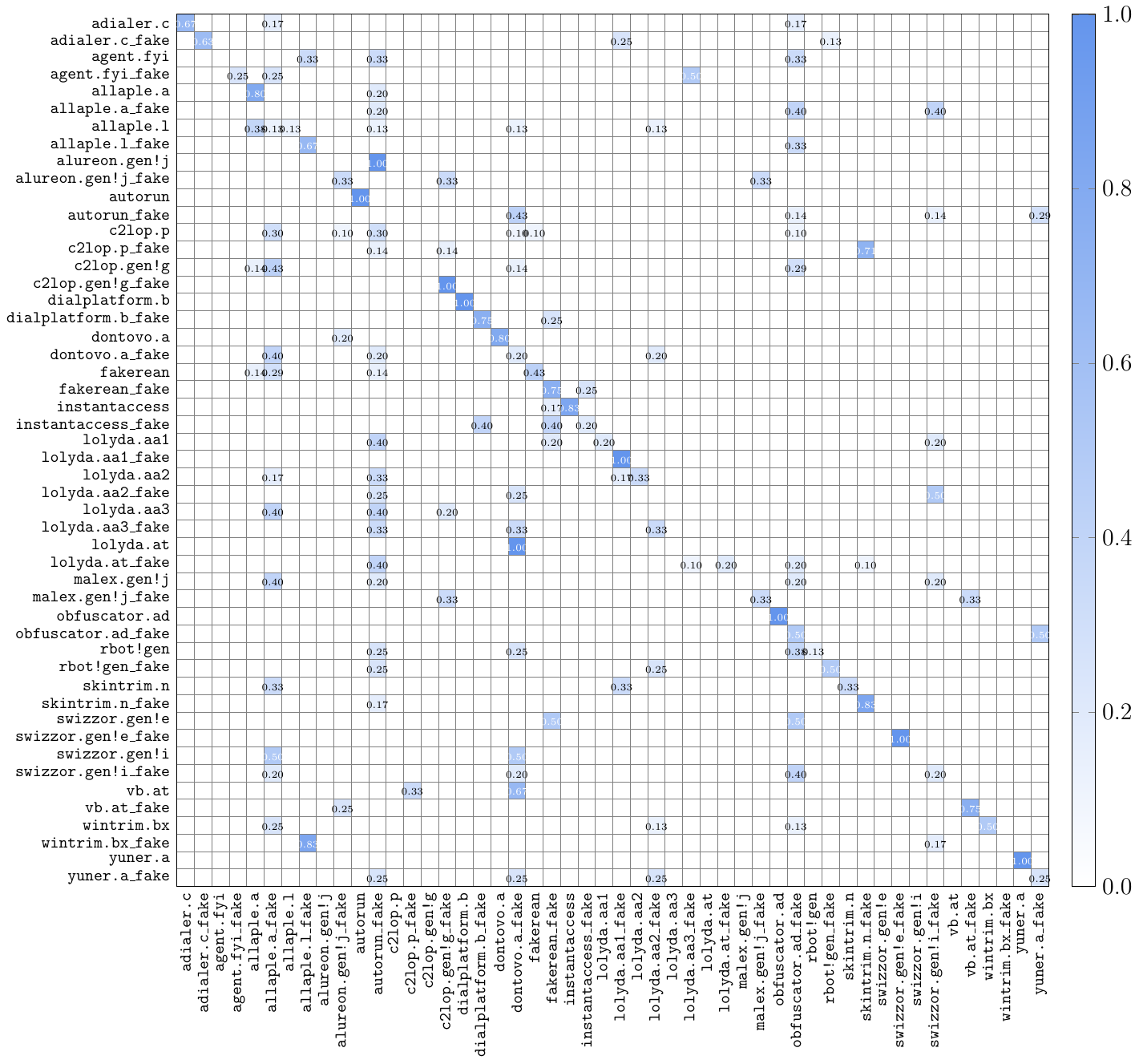}
	\caption{ELM confusion matrix (MalImg $32\times 32$)} 
	\label{fig:app32imgELM}
\end{figure}


\begin{figure}[!htb]
	\centering
	\includegraphics[width=0.85\textwidth]{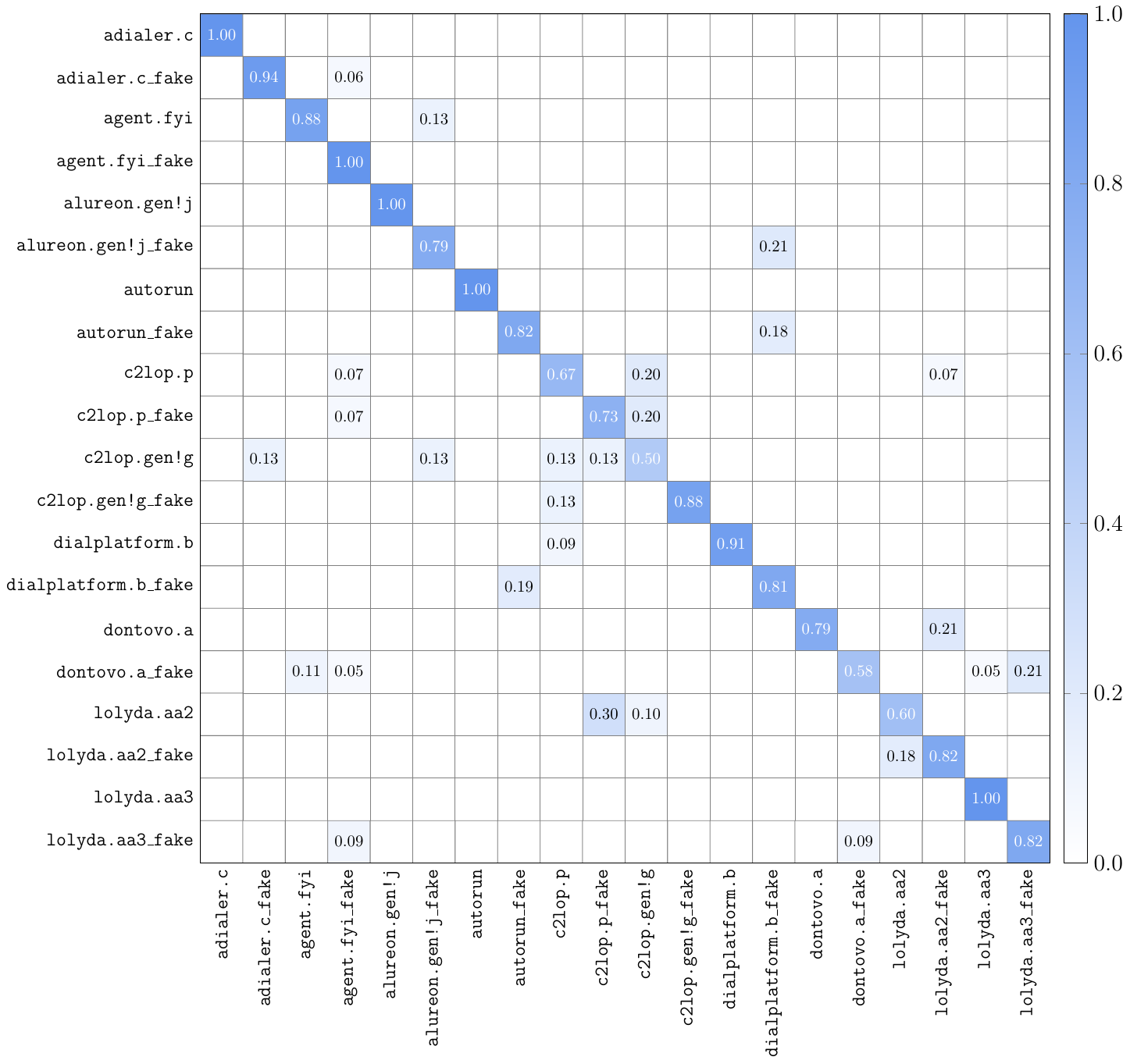}
	\caption{CNN confusion matrix (MalImg $64\times 64$)} 
	\label{fig:app64imgCNN}
\end{figure}

\begin{figure}[!htb]
	\centering
	\includegraphics[width=0.85\textwidth]{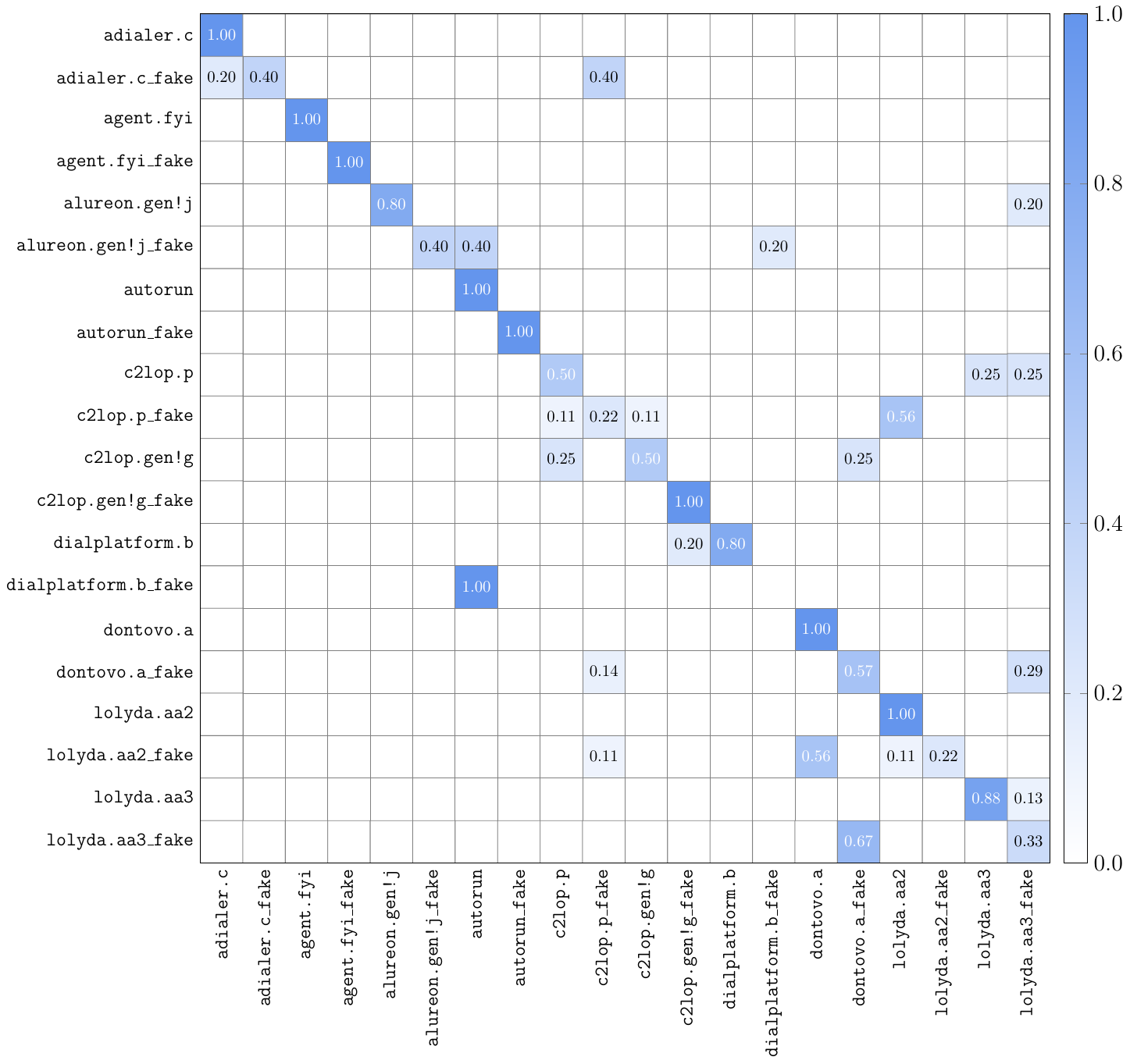}
	\caption{ELM confusion matrix (MalImg $64\times 64$)} 
	\label{fig:app64imgELM}
\end{figure}

\begin{figure}[!htb]
	\centering
	\includegraphics[width=0.85\textwidth]{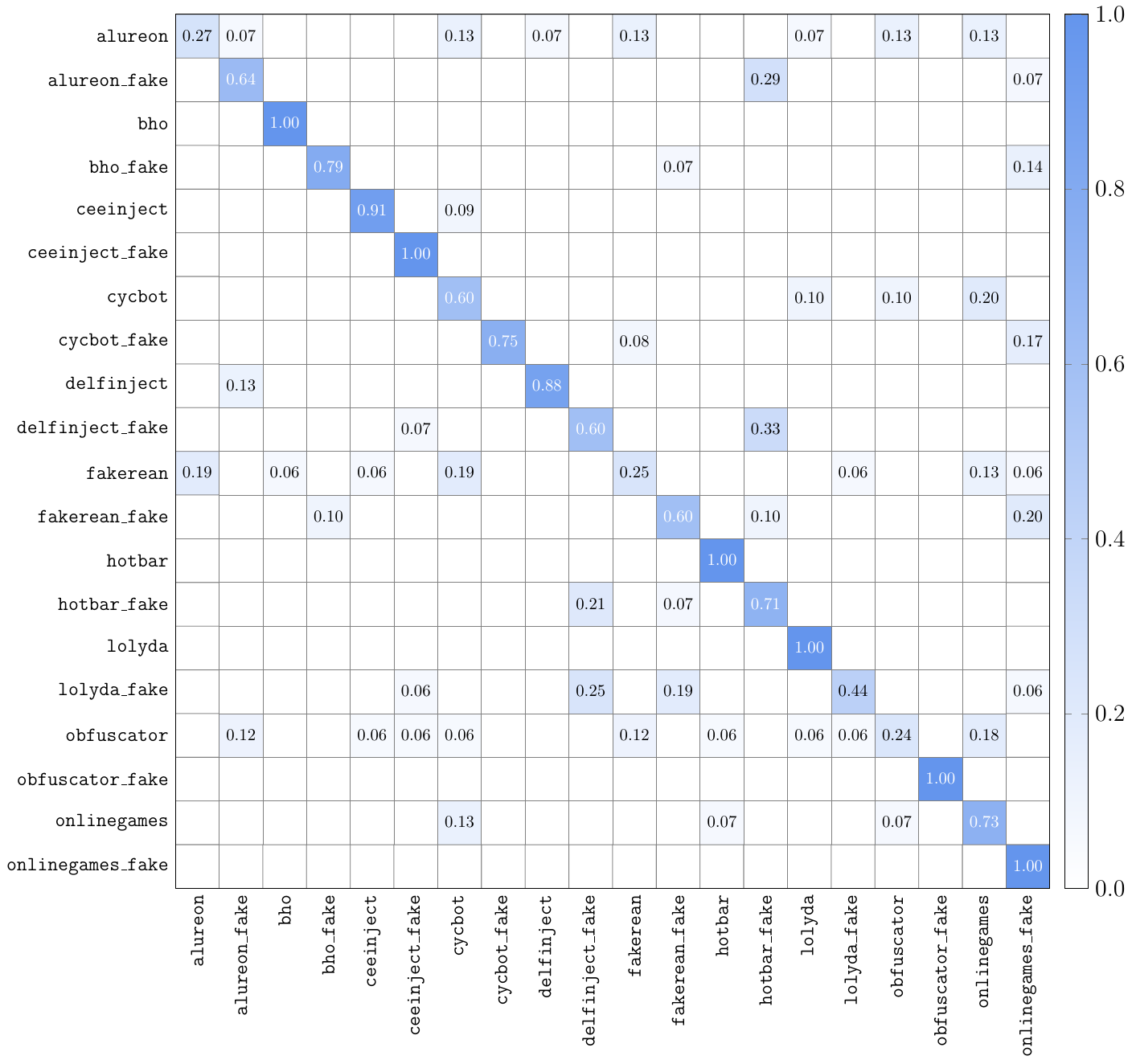}
	\caption{CNN confusion matrix (MalExe $64\times 64$)} 
	\label{fig:app64exeCNN}
\end{figure}

\begin{figure}[!htb]
	\centering
	\includegraphics[width=0.85\textwidth]{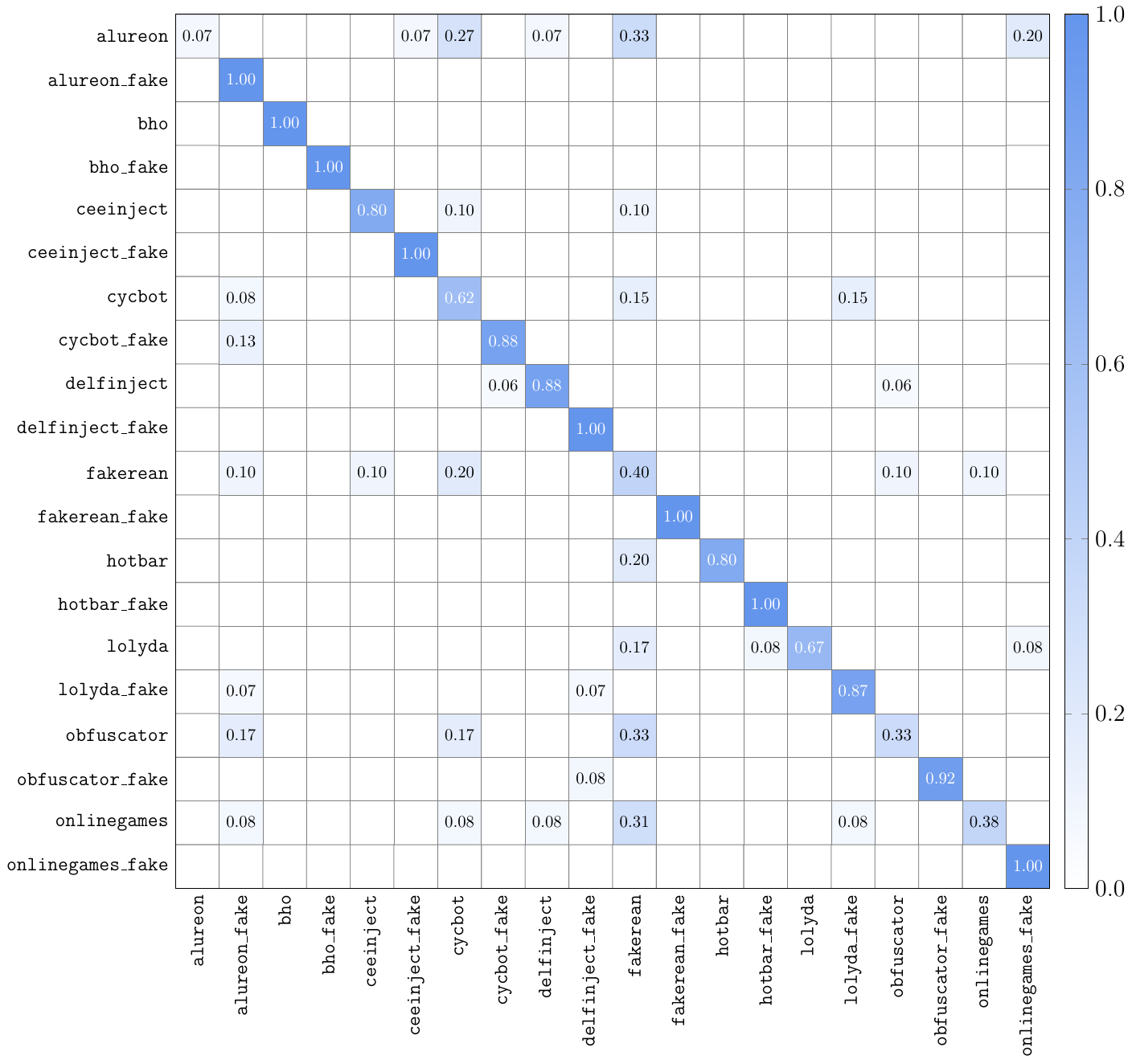}
	\caption{ELM confusion matrix (MalExe $64\times 64$)} 
	\label{fig:app64exeELM}
\end{figure}


\begin{figure}[!htb]
	\centering
	\includegraphics[width=0.85\textwidth]{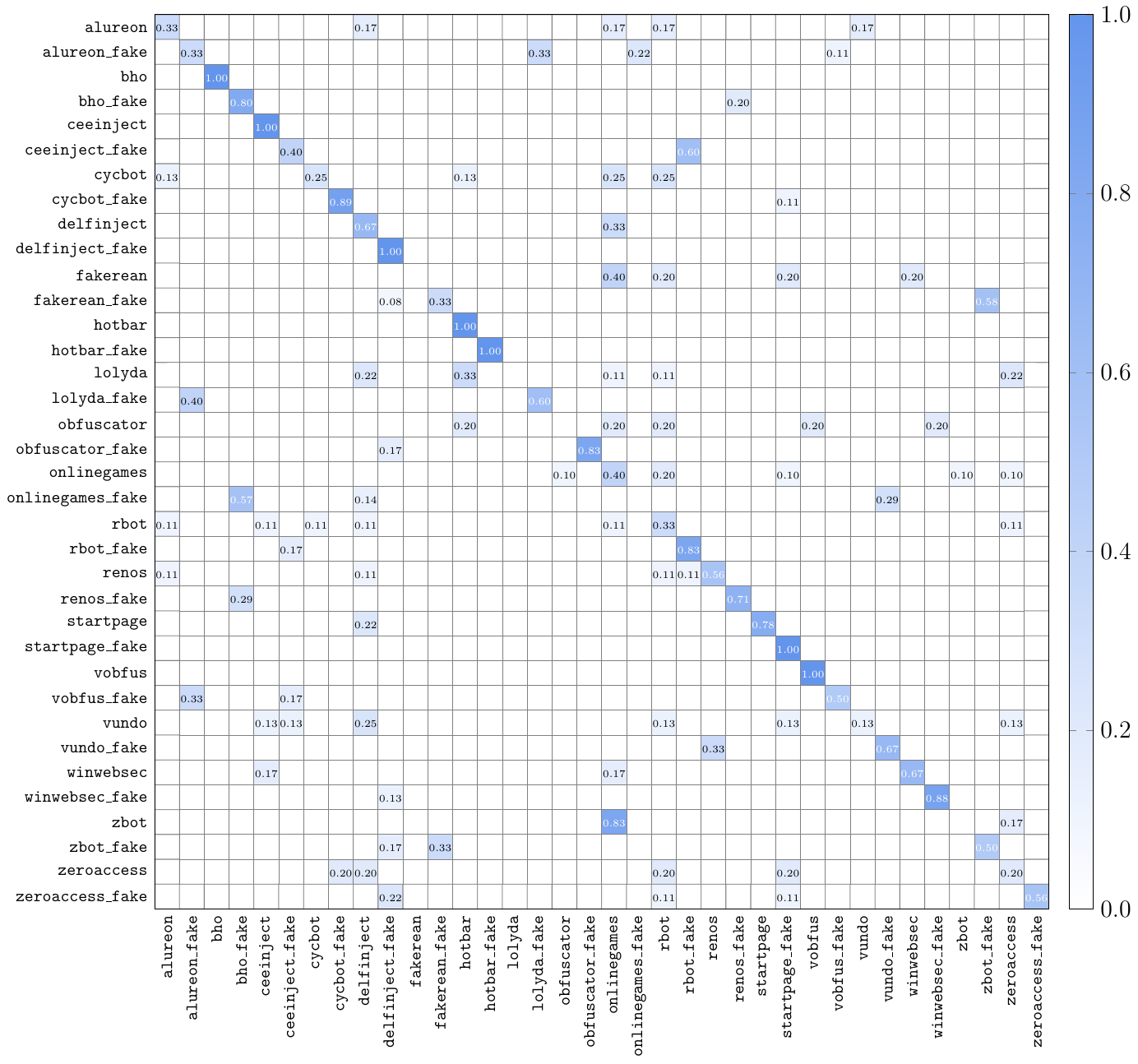}
	\caption{CNN confusion matrix (MalExe $128\times 128$)} 
	\label{fig:app128exeCNN}
\end{figure}

\begin{figure}[!htb]
	\centering
	\includegraphics[width=0.85\textwidth]{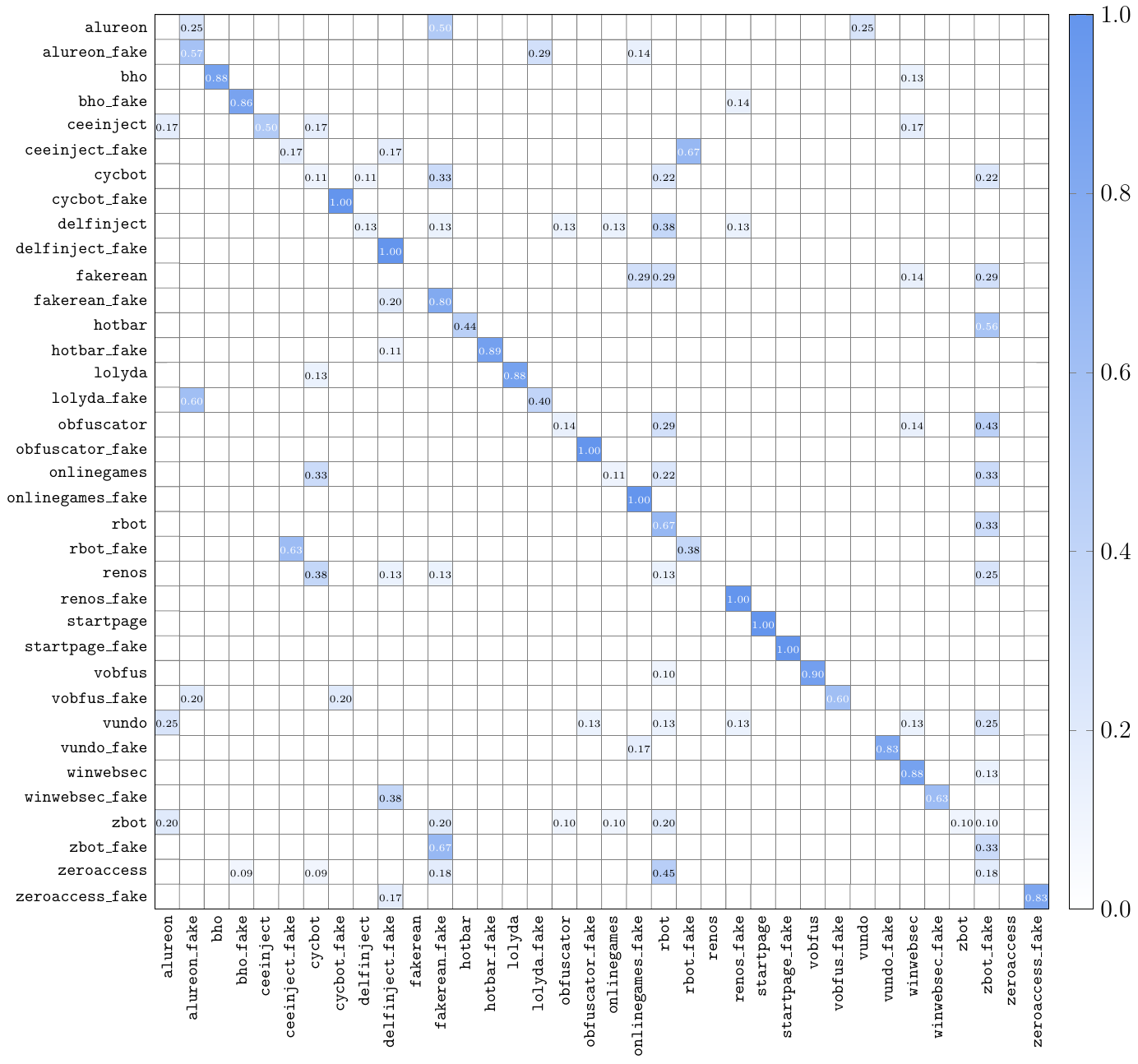}
	\caption{ELM confusion matrix (MalExe $128\times 128$)} 
	\label{fig:app128exeELM}
\end{figure}

\begin{figure}[!htb]
	\centering
	\includegraphics[width=0.85\textwidth]{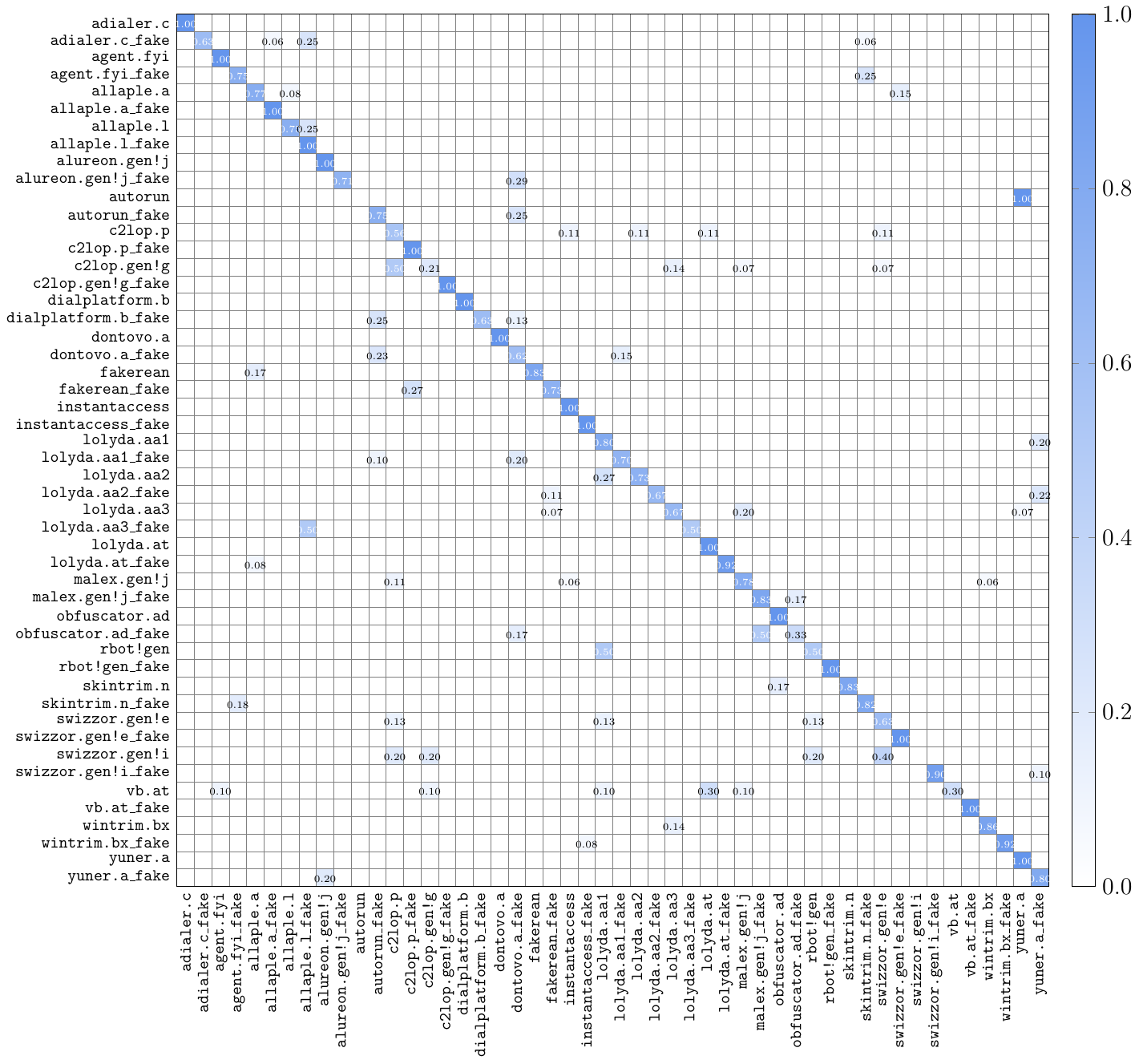}
	\caption{CNN confusion matrix (MalImg $128\times 128$)} 
	\label{fig:app128imgCNN}
\end{figure}

\begin{figure}[!htb]
	\centering
	\includegraphics[width=0.85\textwidth]{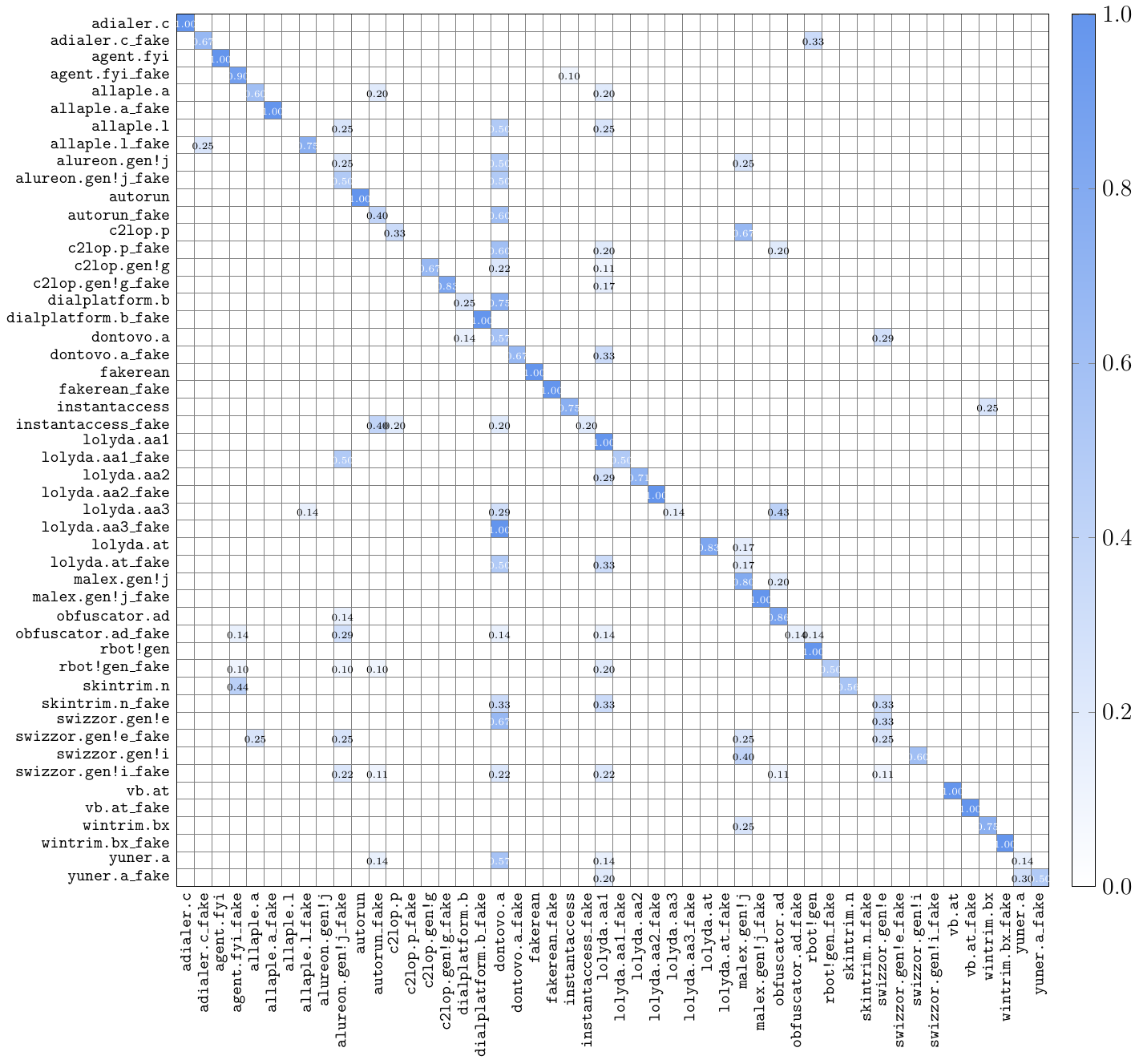}
	\caption{ELM confusion matrix (MalImg $128\times 128$)} 
	\label{fig:app128imgELM}
\end{figure}

\begin{figure}[!htb]
	\centering
	\begin{tabular}{cc}
	\includegraphics[width=0.425\textwidth]{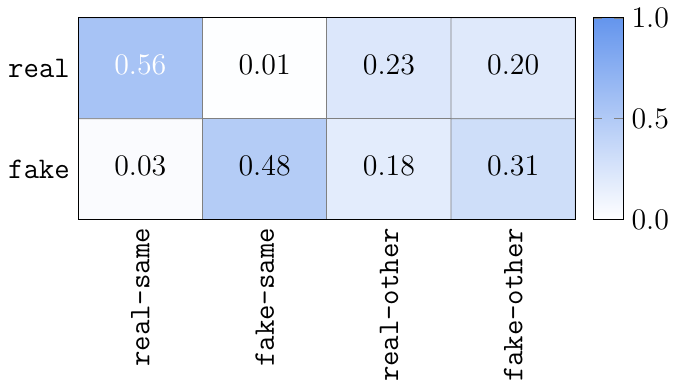}
	& 
	\includegraphics[width=0.425\textwidth]{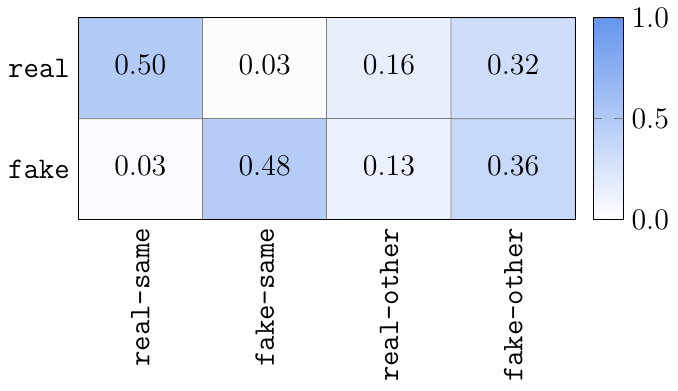}
	\\
	(a) CNN MalExe
	& 
	(b) ELM MalExe
	\\
	\\
	\includegraphics[width=0.425\textwidth]{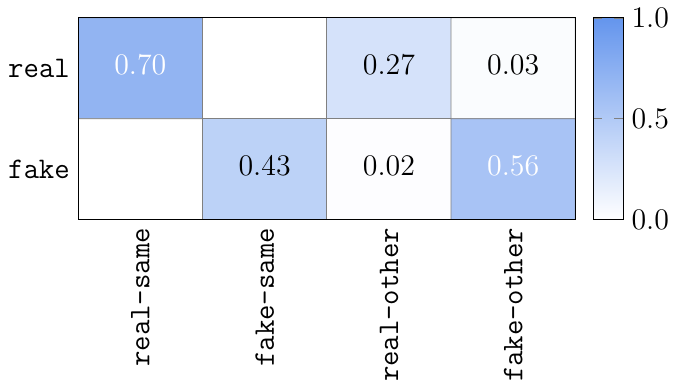}
	& 
	\includegraphics[width=0.425\textwidth]{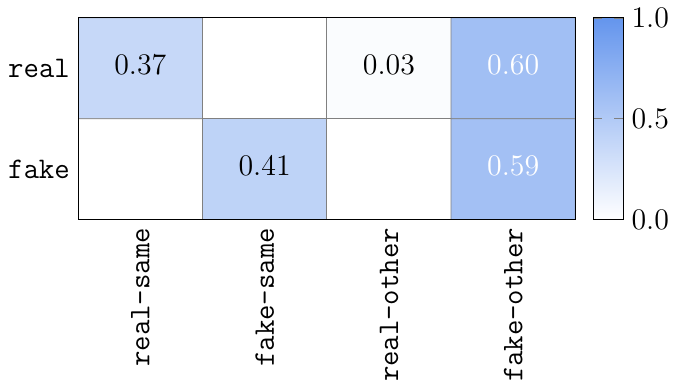}
	\\
	(a) CNN MalImg
	& 
	(b) ELM MalImg
	\end{tabular}
	\caption{Condensed confusion matrices ($32\times 32$)} 
	\label{fig:app32condense}
\end{figure}


\begin{figure}[!htb]
	\centering
	\begin{tabular}{cc}
	\includegraphics[width=0.425\textwidth]{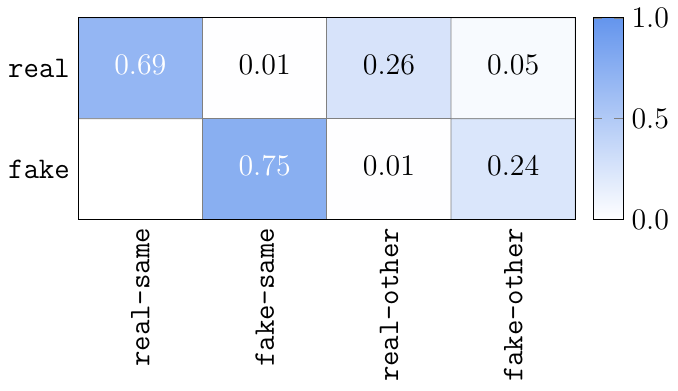}
	& 
	\includegraphics[width=0.425\textwidth]{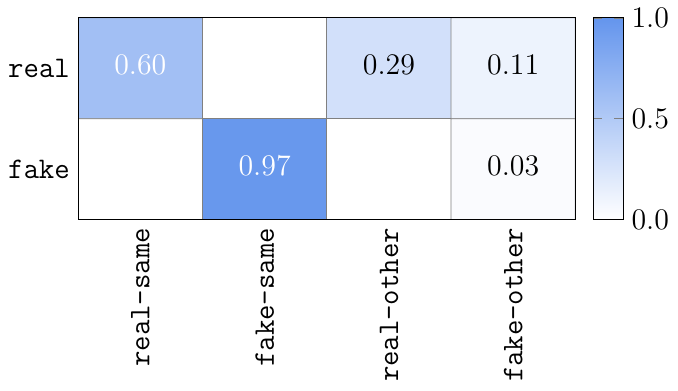}
	\\
	(a) CNN MalExe
	& 
	(b) ELM MalExe
	\\
	\\
	\includegraphics[width=0.425\textwidth]{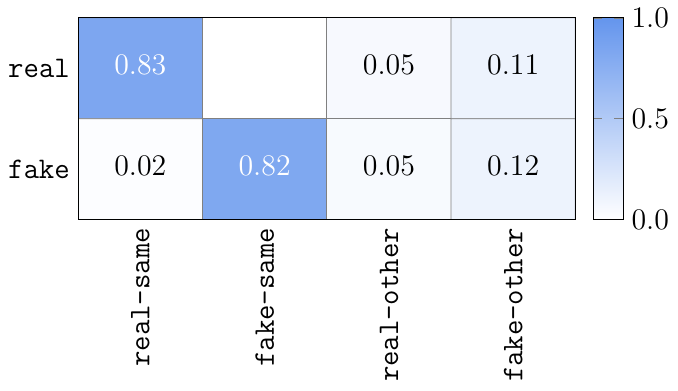}
	& 
	\includegraphics[width=0.425\textwidth]{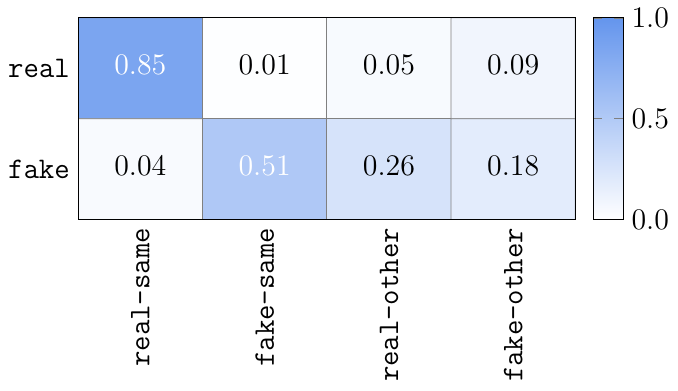}
	\\
	(a) CNN MalImg
	& 
	(b) ELM MalImg
	\end{tabular}
	\caption{Condensed confusion matrices ($64\times 64$)} 
	\label{fig:app64condense}
\end{figure}


\begin{figure}[!htb]
	\centering
	\begin{tabular}{cc}
	\includegraphics[width=0.425\textwidth]{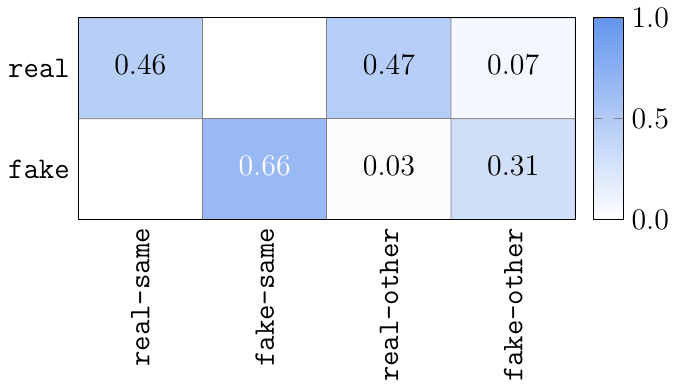}
	& 
	\includegraphics[width=0.425\textwidth]{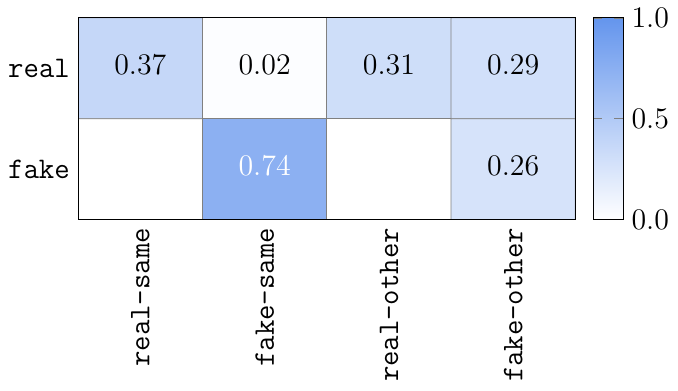}
	\\
	(a) CNN MalExe
	& 
	(b) ELM MalExe
	\\
	\\
	\includegraphics[width=0.425\textwidth]{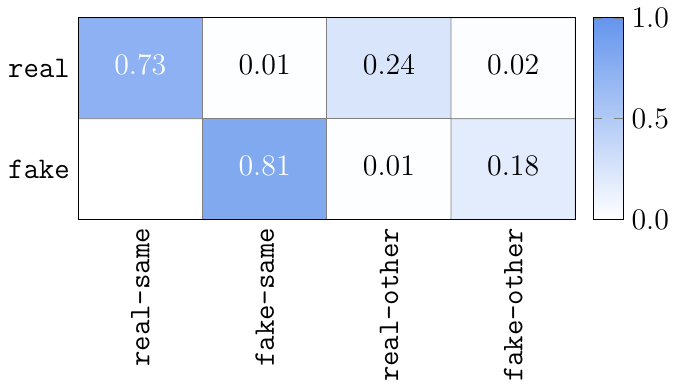}
	& 
	\includegraphics[width=0.425\textwidth]{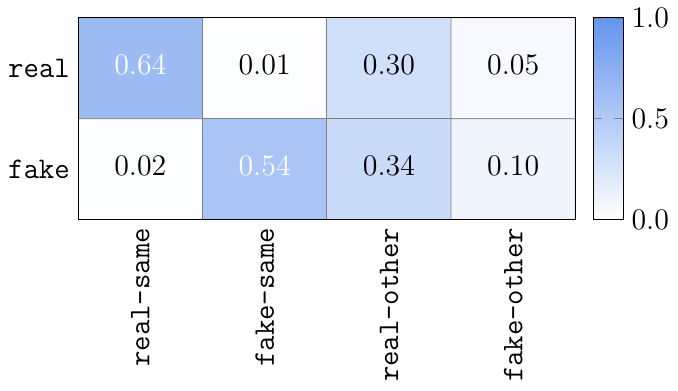}
	\\
	(a) CNN MalImg
	& 
	(b) ELM MalImg
	\end{tabular}
	\caption{Condensed confusion matrices ($128\times 128$)} 
	\label{fig:app128condense}
\end{figure}

\end{document}